\documentclass[12pt]{article}

\usepackage{verbatim}
\usepackage{amsmath}
\usepackage{amssymb}
\usepackage{graphicx}
\usepackage{cite}
\usepackage{subfig}
\usepackage{setspace}
\usepackage{url}
\usepackage[percent]{overpic}
\usepackage{slashed}
\usepackage{authblk}

\usepackage{xspace}
\newcommand{\Fermi}[0]{\textit{Fermi}\xspace}

\newcommand{\sigv}[0]{\ensuremath{\langle\sigma v\rangle}\xspace}
\newcommand{\sigvm}[0]{\ensuremath{\langle\sigma v\rangle_{\rm{UL}}}\xspace}

\newcommand{\unit}[1]{\ensuremath{\mathrm{\,#1}}\xspace}

\newcommand{\GeV}{\unit{GeV}}
\newcommand{\MeV}{\unit{MeV}}
\newcommand{\degree}{\unit{^{\circ}}}

\usepackage{fullpage}
\usepackage{hyperref}
\def\tev{\,{\rm TeV}}
\def\gev{\,{\rm GeV}}

\def\to{\rightarrow}
\def\Fermi{\,{\it Fermi}}

\title{Complementarity and Searches for Dark Matter in the pMSSM}
\date{}
\author[1]{M. Cahill-Rowley}
\author[2]{R. Cotta}
\author[1]{A. Drlica-Wagner}
\author[1]{S. Funk}
\author[1]{J. Hewett}
\author[1]{A. Ismail}
\author[1]{T. Rizzo}
\author[1]{M. Wood}
\affil[1]{SLAC National Accelerator Laboratory, Menlo Park, CA, USA\footnote{mrowley, kadrlica, funk, hewett, aismail, rizzo, mdwood@slac.stanford.edu}}
\affil[2]{University of California, Irvine, CA, USA\footnote{cottar@uci.edu}}

\begin{document}

\rightline{\vbox{\halign{&#\hfil\cr
&SLAC-PUB-15450\cr
&UCI-HEP-TR-2013-09\cr
}}}


{\let\newpage\relax\maketitle}

\begin{abstract}
The search for and identification of neutralino dark matter in supersymmetry requires a multi-pronged approach with important roles played 
by collider, direct and indirect dark matter detection experiments. In this report, we summarize the sensitivity of such searches at the 7, 8 (and 
eventually 14) TeV LHC, combined with those by \Fermi, CTA, IceCube/DeepCore, COUPP and XENON1T, to such particles within the context of the 19-parameter p(henomenological)MSSM. This report provides an outline of the current status 
of our results and our expectations for future analyses. 
\end{abstract}

\section{Introduction and Overview of the pMSSM}

One of the main motivations for R-parity conserving supersymmetry (SUSY) is the prediction that the lightest SUSY particle (LSP) is stable and may be 
identified as a candidate thermal dark matter (DM) particle if it is both electrically neutral and colorless. Quite commonly the role of the LSP is played by 
the lightest neutralino, $\chi_1^0$, and this be will assumed in the discussion here. While DM searches are superficially
focused on the nature of the LSP, the properties of all the other superparticles as well as those of the extended SUSY Higgs sector also come into 
play. Thus it is impossible to completely separate DM searches from searches for and the examination of the rest of the 
SUSY spectrum. However, even in the simplest SUSY scenario, the MSSM, the number of free parameters ($\sim$ 100) is too large to study 
in all generality. 
The traditional approach is to assume the existence of a high-scale theory with only a few parameters (such as mSUGRA{\cite {SUSYrefs}) from 
which all the properties of the sparticles at the TeV scale can be determined and studied in detail. While such an approach is quite valuable~\cite{Cohen:2013kna}, these 
scenarios are somewhat phenomenologically limiting and are under increasing tension with a wide range of experimental data including, in some cases, the $\sim 126$ GeV mass of the recently discovered Higgs boson{\cite {ATLASH,CMSH}}.

One way to circumvent such limitations is to examine the far more general 19-parameter pMSSM{\cite{Djouadi:1998di}}.  The pMSSM is the most general version of the R-parity conserving MSSM when it is subjected to several 
experimentally-motivated constraints: ($i$) CP conservation, ($ii$) Minimal Flavor Violation at the electroweak scale, ($iii$) degenerate first and 
second generation sfermion masses, ($iv$) negligible Yukawa couplings and A-terms for the first two generations. In particular, no assumptions are 
made about physics at high scales, e.g., the nature of SUSY breaking, in order to capture electroweak scale phenomenology for which a UV-complete 
theory may not yet exist. Imposing the constraints ($i$)-($iv$) decreases the number of free parameters in the MSSM at the TeV-scale from 105 to 19 
for the case of a neutralino LSP, or 20 when the gravitino mass is included as an additional parameter when it plays the role of the LSP{\footnote 
{In this work we will limit our discussion to the case of neutralino LSPs}}.  We have 
recently begun a detailed study 
of the signatures for the pMSSM at the 7 and 8 (and eventually 14) TeV LHC, supplemented by input from DM experiments as well as from precision electroweak and 
flavor measurements{\cite {us1,us2,Cahill-Rowley:2013yla}}. In this work, we study and compare the sensitivities of present and future experiments employing different approaches to searching for dark matter: collider production, direct and indirect detection. Independently of the LSP type we will not assume that the thermal relic density 
as calculated for the LSP necessarily saturates the WMAP/Planck value{\cite{Komatsu:2010fb}} to allow for the possibility of multi-component DM. For example, 
the axions introduced to solve the strong CP problem may may make up a substantial amount of DM. The 
19 pMSSM parameters and the ranges of values employed in our scans are listed in Table~\ref{ScanRanges}. The lower and upper limits employed in our scan ranges were chosen to be essentially consistent with Tevatron and LEP data and to have kinematically accessible sparticles at the LHC, respectively. To study the pMSSM, 
we generate many millions of model points in this space (using SOFTSUSY{\cite{Allanach:2001kg}} and checking for consistency using 
SuSpect{\cite{Djouadi:2002ze}}), with each point corresponding to a specific set of values for these parameters. 
These individual models are then subjected to a large set of collider, flavor, precision measurement, dark matter and theoretical constraints~\cite{us1}.  
Roughly 225k models with a neutralino LSP survive this initial selection and can then be used for further physics studies. (We have also  
obtained a model set of similar size for the case of gravitino LSPs.) Decay patterns of the SUSY partners and the extended Higgs sector 
are calculated using a modified version of SUSY-HIT{\cite{Djouadi:2006bz}}. 

In addition to these two large pMSSM model sets, we have recently generated a smaller, specialized neutralino LSP set of $\sim$ 10k `natural' models 
all of which predict $m_h=126\pm 3$ GeV, have an LSP that {\it does} saturate the WMAP relic density and which produce values of fine-tuning (FT) 
better than $1\%$ using the Barbieri-Giudice measure~\cite{Ellis:1986yg, Barbieri:1987fn}. This low-FT model set will also be used in the future as part of our completed study.

\begin{table}
\centering
\begin{tabular}{|c|c|} \hline\hline
$m_{\tilde L(e)_{1,2,3}}$ & $100 \gev - 4 \tev$ \\ 
$m_{\tilde Q(q)_{1,2}}$ & $400 \gev - 4 \tev$ \\ 
$m_{\tilde Q(q)_{3}}$ &  $200 \gev - 4 \tev$ \\
$|M_1|$ & $50 \gev - 4 \tev$ \\
$|M_2|$ & $100 \gev - 4 \tev$ \\
$|\mu|$ & $100 \gev - 4 \tev$ \\ 
$M_3$ & $400 \gev - 4 \tev$ \\ 
$|A_{t,b,\tau}|$ & $0 \gev - 4 \tev$ \\ 
$M_A$ & $100 \gev - 4 \tev$ \\ 
$\tan \beta$ & 1 - 60 \\
$m_{3/2}$ & 1 eV$ - 1 \tev$ ($\tilde{G}$ LSP)\\
\hline\hline
\end{tabular}
\caption{Scan ranges for the 19 (20) parameters of the pMSSM with a neutralino (gravitino) LSP. The gravitino mass is scanned with a log prior. 
All other parameters are scanned with flat priors, though we expect this choice to have little qualitative impact on our results~\cite{us}.}
\label{ScanRanges}
\end{table}

As a result of the upper limit employed in our scan ranges of 4 TeV, an upper limit chosen to enable phenomenological studies at the 14 TeV LHC, the LSPs in our model sets are typically very close to being a pure 
electroweak eigenstate since the off-diagonal elements of the chargino and neutralino mass matrices are at most $\sim M_W$. Figure~\ref{fig0} 
shows some properties of the nearly pure neutralino LSPs (defined as a single electroweak eigenstate comprising over 90\% of the mass eigenstate). In the left panel we see the distribution of the LSP mass for nearly pure bino, wino, and Higgsino LSPs while in the right-hand panel we see the corresponding distribution for the predicted LSP thermal relic 
density. Note that the masses of all of our neutralino LSPs lie below $\sim 2$ TeV due to our choice of the scan ranges, and the entire SUSY 
spectrum must be heavier than the LSP and less than $\sim 4$ TeV by definition. 
The fraction of models which are nearly pure bino is found to be rather low in this model set since pure binos generally predict too high 
a relic density unless they co-annihilate with another sparticle, happen to be close to some ($Z,h,A$) funnel region or have a suitable Higgsino admixture. Note 
that only in the rightmost bin of the right panel is the relic density approximately saturating the WMAP thermal relic value. These LSP properties will be of particular importance 
in the discussion that follows.

\begin{figure}[htbp]
\centerline{\includegraphics[width=3.5in]{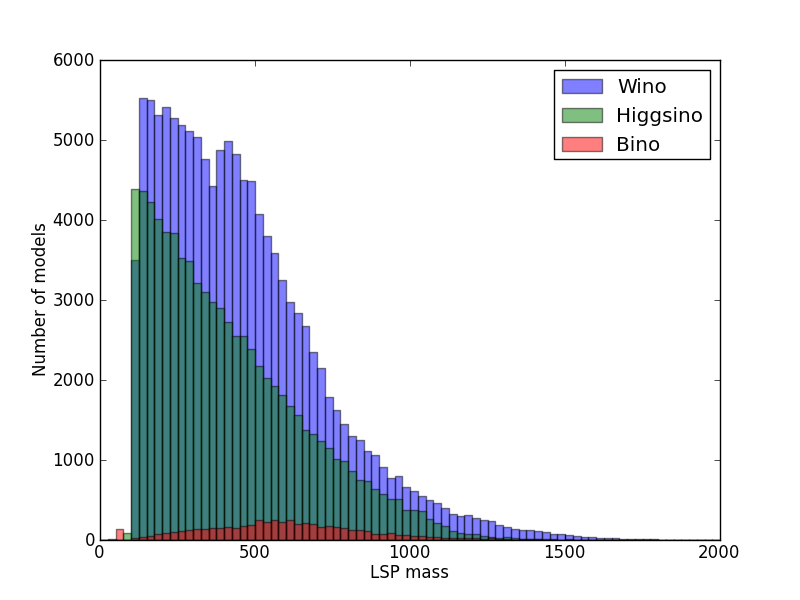}
\hspace{-0.50cm}
\includegraphics[width=3.5in]{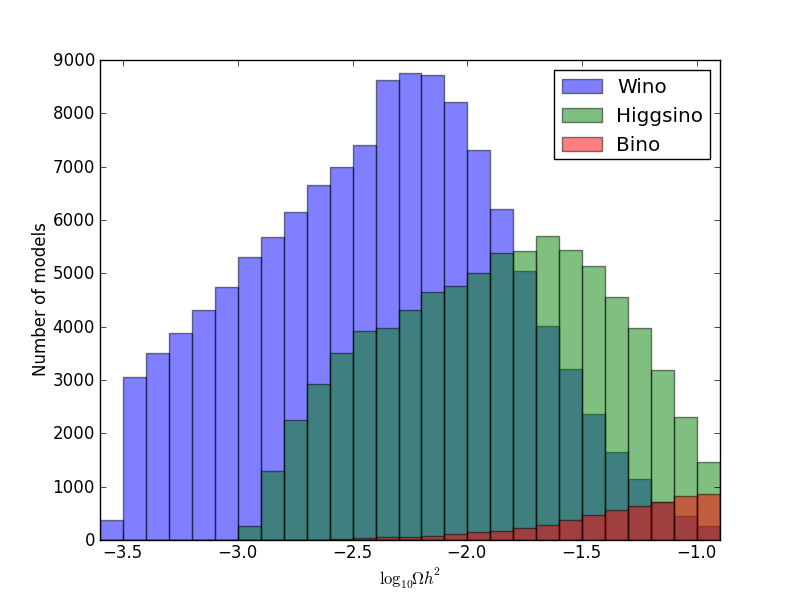}}
\vspace*{-0.10cm}
\caption{Distribution of the LSP masses (left) and predicted relic densities (right) for neutralino LSPs which are almost pure weak eigenstates.}
\label{fig0}
\end{figure}

Figure~\ref{fig00} shows the thermal relic densities of the LSPs in our pMSSM model sample as a function of their mass with the color-coding reflecting their electroweak 
eigenstate content. There are many things to note here that will be important for later consideration; essentially every possible mechanism to 
obtain (or lie below) the WMAP relic density is present: ($i$) The set of models at low masses on the left-hand side forming `columns' correspond 
to bino and bino-Higgsino admixtures surviving due to their proximity to the $Z,h$-funnels. ($ii$) The bino-Higgsino LSPs saturating the relic density 
in the upper left are of the so-called 'well-tempered' variety. ($iii$) the pure{\footnote {Here again, `pure' means having an eigenstate fraction 
$\geq 90\%$. Points shown as bino-wino, bino-Higgsino, or wino-Higgsino mixtures have less than $2\%$ Higgsino, wino, or bino fraction, respectively. Mixed points have 
no more than $90\%$ and no less than $2\%$ of each component.}} bino models in the middle top of the Figure are bino co-annihilators (mostly with 
sleptons) or  are models near the $A$-funnel region. ($iv$) The green (blue) bands are pure 
Higgsino (wino) models that saturate the relic density bound near $\sim 1(1.7)$ TeV but dominantly appear at far lower relic densities. Wino-Higgsino 
hybrids are seen to lie between these two cases as expected. ($v$) A smattering of other annihilator models are seen to be loosely distributed in the 
lower right-hand corner of the figure.

\begin{figure}[htbp]
\centerline{\includegraphics[width=7.0in]{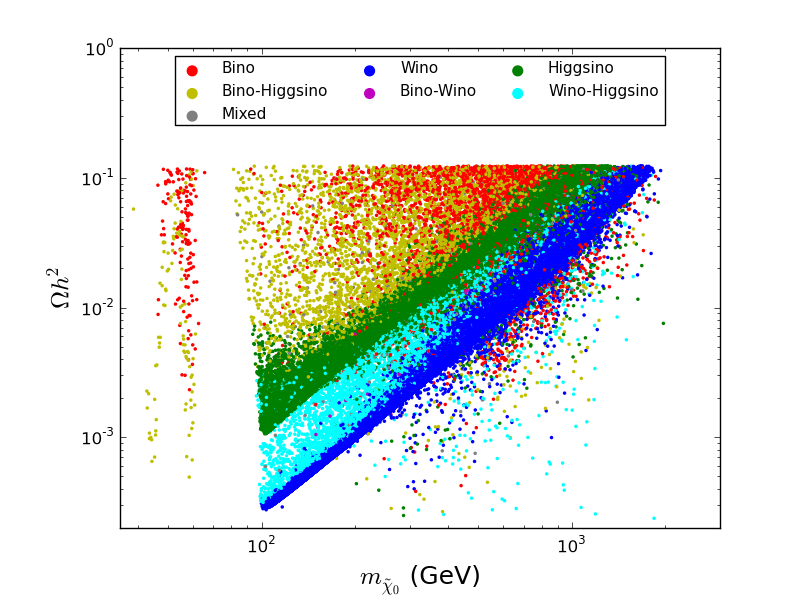}}
\vspace*{-0.10cm}
\caption{Thermal relic density as a function of the LSP mass in our pMSSM model set, as generated, color-coded by the electroweak properties of the  
LSP as discussed in the text.}
\label{fig00}
\end{figure}

\section{LHC Searches}

We begin with a short overview of the searches for the pMSSM at the 7 and 8 TeV LHC~\cite{us1,us2,Cahill-Rowley:2013yla}. In general, we follow the suite of ATLAS SUSY 
analyses but also supplement these with several searches performed by CMS; the analyses are briefly summarized in Table~\ref{SearchList}. The SUSY 
searches are then further augmented by additional searches for heavy neutral SUSY Higgs $\to \tau^+\tau^-$ by CMS~\cite{Chatrchyan:2012vp} and for the rare decay mode  
$B_s\to \mu^+\mu^-$ as discovered by CMS and LHCb~\cite{BSMUMU}, both of which play distinct but important roles in restricting the pMSSM parameter space. Presently, we have 
implemented every relevant ATLAS SUSY search publicly available as of the beginning of March 2013. This list is currently being 
updated and expanded for future analysis. The LHC analysis results for all three of our model sets (including the neutralino LSP model set considered in this paper) appear in detail in our companion HE4 Snowmass 
White Paper on SUSY searches~\cite{Cahill-Rowley:2013yla}. 

Briefly stated our procedure is as follows: We generate SUSY events for each model using PYTHIA 6.4.26~\cite{Sjostrand:2006za} and PGS 4~\cite{PGS}, 
which we have modified to, e.g., correctly deal with gravitinos, hadronization of stable colored sparticles, multi-body decays and ATLAS b-tagging. 
We then scale our event rates to NLO by calculating the relevant K-factors using Prospino 2.1~\cite{Beenakker:1996ch}. The individual searches are 
then implemented using a custom analysis code{\cite {us}}, following the published cuts and selection criteria as closely as possible. 
This analysis 
code is validated for each of the many search regions in every analysis employing the benchmark model points provided by ATLAS (and CMS). Models are then 
excluded using the 
same $CL_s$ criterion as is used by ATLAS. These analyses are performed {\it without} imposing the Higgs mass constraint, $m_h=126 \pm 3$ GeV (combined 
experimental and theoretical errors) so that we can understand its influence on the search results. Note that roughly $20\%$ of the neutralino model 
set (before the LHC SUSY searches are applied) predict a Higgs mass in the above range. While there is some variation amongst the individual searches we 
find that, once combined, the total fraction of our models surviving (or killed by) the set of all LHC searches is to an excellent approximation 
{\it independent} of whether or not the Higgs mass constraint has been applied. Conversely, the $\sim 20\%$ fraction of neutralino models predicting the correct 
Higgs mass is also found to be approximately independent of whether or not the SUSY searches have been applied. This combined result is 
very powerful, demonstrating the current approximate decoupling of SUSY search results from the discovery of the Higgs boson and allowing us to continue 
examining the entire model set with some validity.

\begin{table}
\centering
\begin{tabular}{|l|l|l|} \hline\hline
Search  & Energy &   Reference     \\
\hline
2-6 jets & 7 TeV & ATLAS-CONF-2012-033   \\
multijets & 7 TeV & ATLAS-CONF-2012-037    \\
1 lepton  & 7 TeV & ATLAS-CONF-2012-041    \\

2-6 jets   & 8 TeV &   ATLAS-CONF-2012-109   \\
multijets   & 8 TeV &  ATLAS-CONF-2012-103    \\
1 lepton     & 8 TeV &  ATLAS-CONF-2012-104   \\
SS dileptons & 8 TeV &  ATLAS-CONF-2012-105    \\

Gluino $\to$ Stop/Sbottom   & 7 TeV & 1207.4686  \\
Very Light Stop  & 7 TeV & ATLAS-CONF-2012-059   \\
Medium Stop  & 7 TeV & ATLAS-CONF-2012-071 \\
Heavy Stop (0l)  & 7 TeV & 1208.1447 \\
Heavy Stop (1l)   & 7 TeV & 1208.2590  \\
GMSB Direct Stop  & 7 TeV & 1204.6736   \\
Direct Sbottom & 7 TeV & ATLAS-CONF-2012-106  \\
3 leptons & 7 TeV & ATLAS-CONF-2012-108  \\
1-2 leptons & 7 TeV & 1208.4688  \\
Direct slepton/gaugino (2l)  & 7 TeV & 1208.2884  \\
Direct gaugino (3l) & 7 TeV & 1208.3144  \\
4 leptons & 7 TeV & 1210.4457  \\
1 lepton + many jets & 7 TeV & ATLAS-CONF-2012-140  \\
1 lepton + $\gamma$ & 7 TeV & ATLAS-CONF-2012-144  \\
$\gamma$ + b & 7 TeV & 1211.1167  \\
$\gamma \gamma $ + MET ($\tilde{G}$ LSP) & 7 TeV & 1209.0753  \\

Medium Stop (2l) & 8 TeV & ATLAS-CONF-2012-167   \\
Medium/Heavy Stop (1l) & 8 TeV & ATLAS-CONF-2012-166  \\
Direct Sbottom (2b) & 8 TeV & ATLAS-CONF-2012-165  \\
3rd Generation Squarks (3b) & 8 TeV & ATLAS-CONF-2012-145  \\
3rd Generation Squarks (3l) & 8 TeV & ATLAS-CONF-2012-151  \\
3 leptons & 8 TeV & ATLAS-CONF-2012-154  \\
4 leptons & 8 TeV & ATLAS-CONF-2012-153  \\
Z + jets + MET & 8 TeV & ATLAS-CONF-2012-152  \\

HSCP      & 7 TeV  &  1205.0272    \\
Disappearing tracks  & 7 TeV  &  ATLAS-CONF-2012-111  \\
Muon + Displaced Vertex ($\tilde{G}$ LSP) & 7 TeV  & 1210.7451  \\
Displaced Dilepton ($\tilde{G}$ LSP) & 7 TeV  & 1211.2472  \\

$B_s \to \mu \mu$ & 7+8 TeV  & 1211.2674  \\
$A/H \to \tau \tau$ & 7+8 TeV  & CMS-PAS-HIG-12-050  \\

\hline\hline
\end{tabular}
\caption{Simulated LHC SUSY searches which have been applied to our pMSSM model sets. 63 (48) percent of our models with a neutralino 
(gravitino) LSP are not excluded by any of these searches and this is found to be approximately independent of the Higgs mass constraint.}
\label{SearchList}
\end{table}

Figure~\ref{fig1} shows some results demonstrating the power of this combination of LHC analyses to constrain the set of pMSSM neutralino LSP 
models. In particular this figure shows the fraction of models with a given sparticle and LSP mass which are excluded by the combined LHC searches. Since the masses 
of the squarks and gluinos, the lightest stops and sbottoms and the LSP itself are currently of particular interest, we have concentrated on those 
quantities in this figure. In the upper left-hand panel the coverage of the gluino-LSP mass plane by the LHC searches is displayed; the solid white 
line represents the Simplified Model $95\%$ CL search limit from ATLAS~\cite{ATLAS:2012ee} which we see is roughly the same as the all-black region excluded in the pMSSM. The upper right-hand panel shows the corresponding coverage in the gluino-lightest 
squark mass plane with the analogous Simplified Model line from ATLAS~\cite{ATLAS:2012ee} (which assumes a zero mass LSP and that the 8 squarks of the first two 
generations are degenerate, neither of which are necessarily true in the pMSSM). While most models with rather light squarks and/or gluinos are observed 
to be excluded by the LHC searches, it is clear that models with both squarks and gluinos below $\sim 700$ GeV still remain viable. The lower 
two panels show the lightest stop/sbottom vs. LSP mass plane with the corresponding ATLAS Simplified Model search limits~\cite{ATLAS3glimits}. Here we see several things,  
the most important being that the amount of coverage of these two mass planes in the pMSSM by the LHC searches is significantly larger than what we 
would have expected from either Simplified Model analysis. We also find that the third generation searches are ineffective for models that have a 
relatively small splitting between the stop and LSP ($\delta m \lesssim$ 130 GeV), so the exclusion of these models relies almost entirely on the 
standard jets + MET searches. In the uncompressed regions, on the other hand, the third generation searches significantly outperform jets + MET, as expected. In particular, 
the $2b$+MET analysis is found to be responsible for nearly all of the exclusion power in this region. Clearly, the third-generation searches cover important gaps in the standard jets + MET coverage. In general, the third-generation searches are most sensitive to models 
in which stop or sbottom decays produce hard $b$-jets, and have a more difficult time observing models with stops or sbottoms that produce tops or 
decay to multiple intermediate gauginos (resulting in softer $b$-jets and leptons). Lastly we again stress that the direct sbottom search is found to be 
responsible for the majority of the heavy flavor search exclusion power for models with relatively light stops.

\begin{figure}[htbp]
\centerline{\includegraphics[width=3.5in]{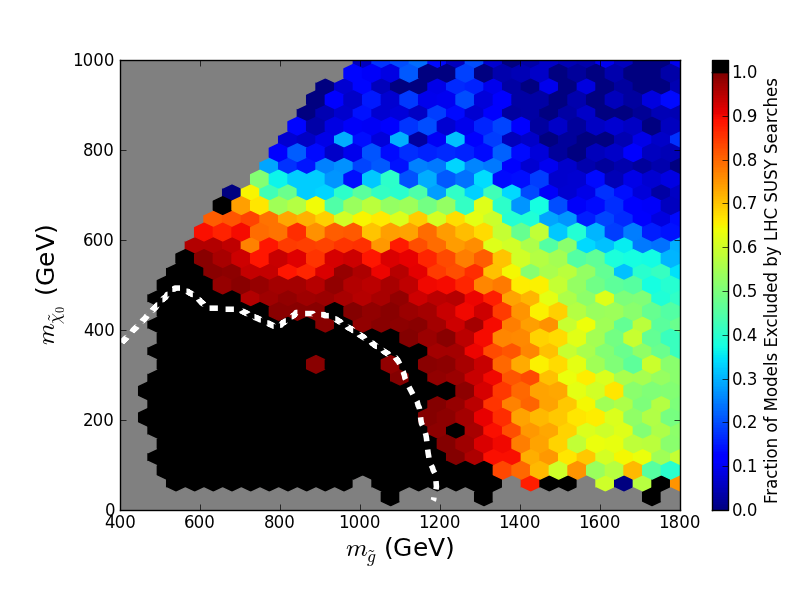}
\hspace{-0.50cm}
\includegraphics[width=3.5in]{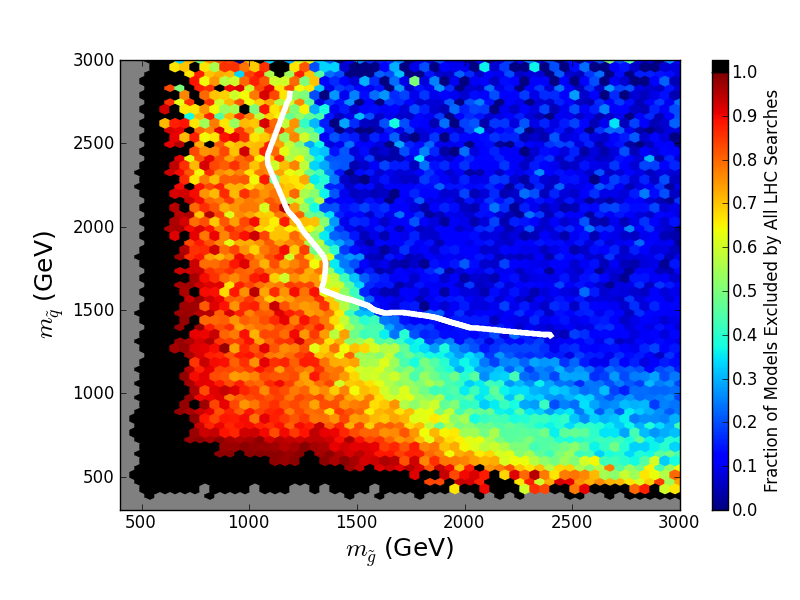}}
\vspace*{0.50cm}
\centerline{\includegraphics[width=3.5in]{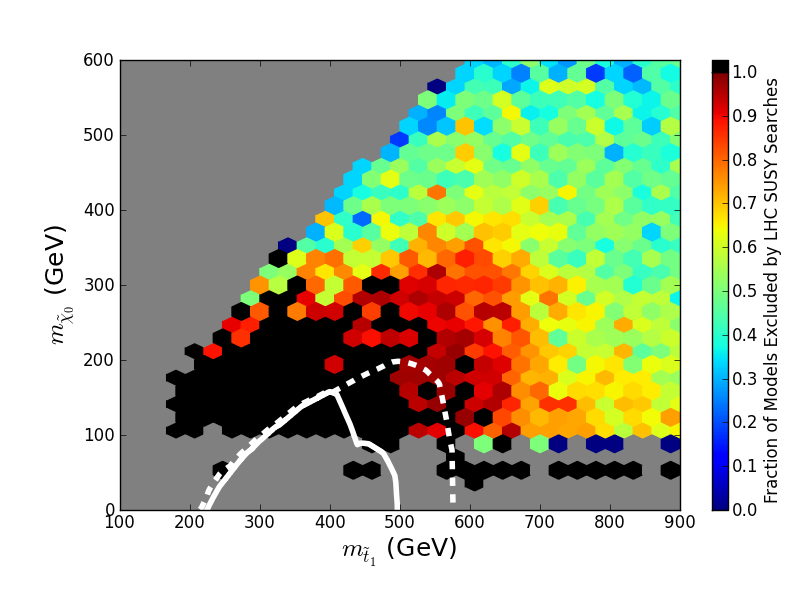}
\hspace{-0.50cm}
\includegraphics[width=3.5in]{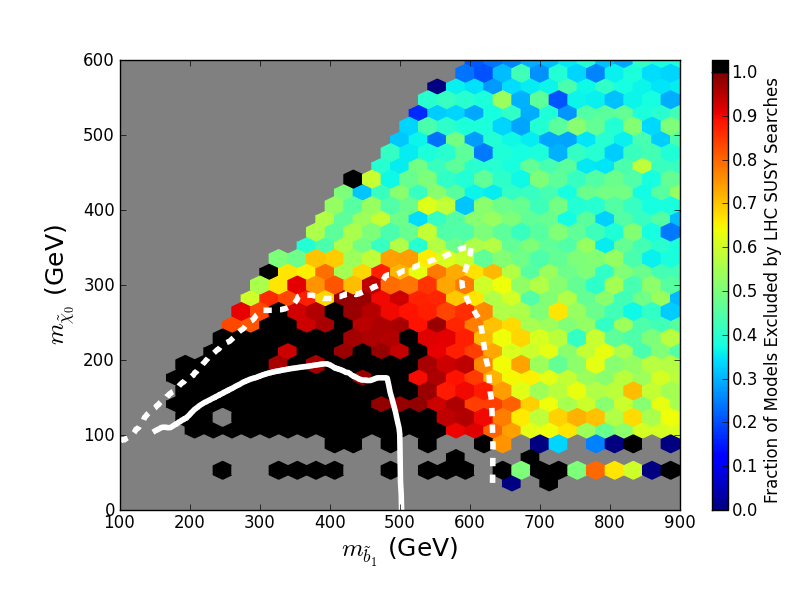}}
\vspace*{-0.10cm}
\caption{Projections of the pMSSM model coverage efficiencies from the 7 and 8 TeV LHC searches shown in the gluino-LSP mass plane (top left), 
the gluino-lightest squark mass plane (top right), the lightest stop-LSP mass plane (lower left) and in the lightest sbottom-LSP mass plane. The 
solid lines represent the corresponding $95\%$ CL limit results obtained by ATLAS in the simplified model limit as discussed in the text.}
\label{fig1}
\end{figure}

In addition to the 7 and 8 TeV LHC searches, future data taking and enhanced analyses at $\sim$ 14 TeV will greatly extend the expected coverage of the 
pMSSM parameter space for both LSP types. In~\cite{Cahill-Rowley:2013yla}, we consider the impact of one of the most powerful of these searches to be performed by ATLAS - the zero-lepton jets + 
MET final state~\cite {ATLAS-EP}} - following the same procedure that was used for the 7/8 TeV analyses. Note that in extrapolating from 300 fb$^{-1}$ to 3 ab$^{-1}$ luminosity scaling has been employed to obtain the expected limit. As a result of limitations on CPU time, we only generated 14 TeV results for the $\sim 30.7$k 
neutralino LSP models that survive the 7 and 8 TeV LHC analyses and also predict a Higgs mass of $126\pm 3$ GeV. We note that since the results of the 7 and 
8 TeV analyses are essentially independent of the Higgs mass, it is quite likely that our results for this narrow Higgs mass range would in fact be applicable, 
at least to a very good approximation, to the entire neutralino LSP model set.

\begin{figure}[htbp]
\centerline{\includegraphics[width=3.5in]{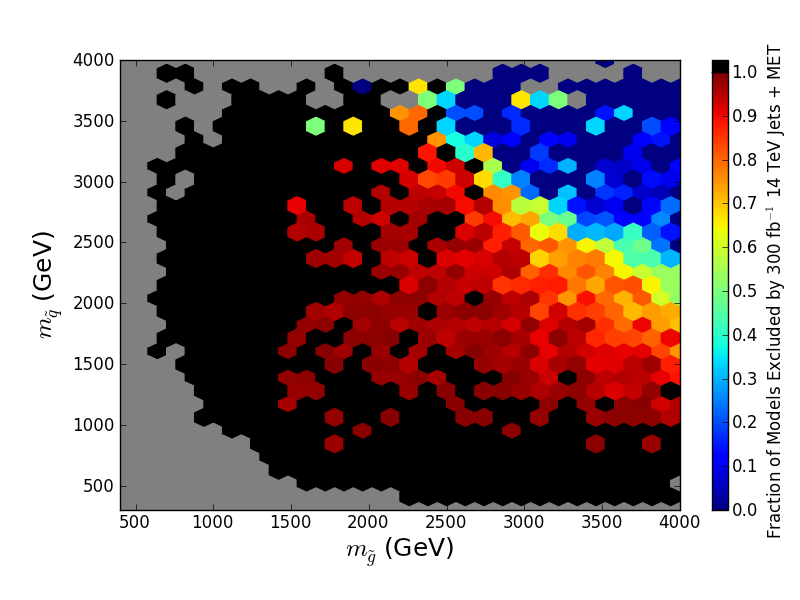}
\hspace{-0.50cm}
\includegraphics[width=3.5in]{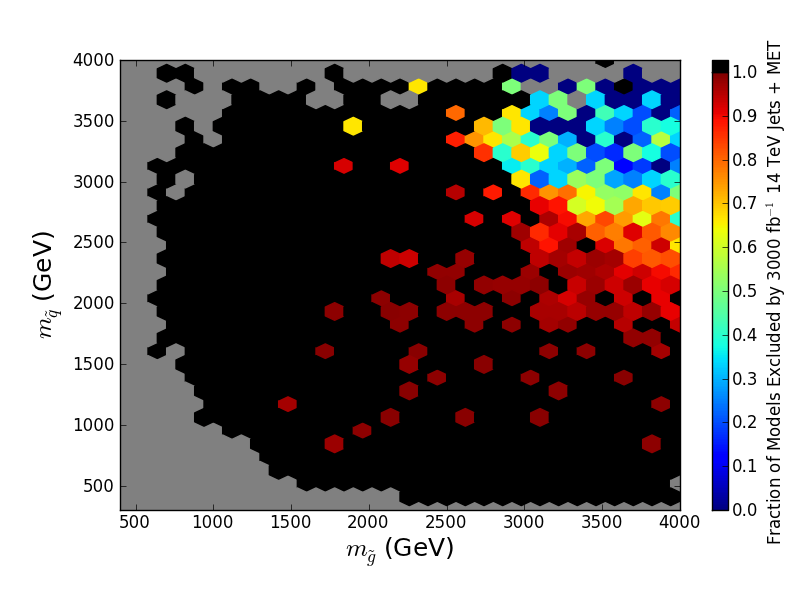}}
\vspace*{-0.10cm}
\caption{Expected results from a jets + MET search at the 14 TeV LHC assuming an integrated luminosity of 300 fb$^{-1}$ (left) and 
3000 fb$^{-1}$ (right), in the lightest squark-gluino mass plane.}\label{fig2}
\end{figure}

Our results for the ATLAS 14 TeV zero-lepton jets+MET search are summarized in 
Fig.~\ref{fig2} for both 0.3 and 3 ab$^{-1}$ of integrated luminosity in the gluino-lightest squark mass plane. Here we see that for typical models, the 14 TeV LHC will be able to exclude squarks up to $\sim$ 1.6 TeV and gluinos up to $\sim$ 2.7 TeV, with only 7.9 (2.5) percent of models surviving after 300 (3000) fb$^{-1}$ of data. A few models are seen to survive with much lighter squark and/or gluino masses; in most cases these models survive by producing high-$p_T$ leptons and therefore falling outside of the 0-lepton search region, so adding other searches with leptons in the signal region will undoubtedly exclude many of the surviving models with light colored sparticles. To 
fully understand the capabilities of the 14 TeV LHC will of course require a far more realistic study than is presently available since the LHC collaborations themselves are 
still unsure of how their detectors will perform at 14 TeV with very high pileup conditions, and we can currently only simulate one of the 14 TeV search channels (although this search is the most powerful at 7/8 TeV and is expected to dominate the 14 TeV exclusions as well).

\section{Direct Detection}

The direct detection of DM results from either the spin-independent (SI) or spin-dependent (SD) scattering of the LSP by a target nucleus. While 
$Z(h)$ $t-$channel exchange only contributes to the SD(SI) process at tree level, $s-$channel squark exchange can contribute to both processes. 
However, as the lower bounds on the first and second generation squark masses from the LHC become stronger the importance of these squark exchange contributions 
clearly become subdominant.{\footnote {It is important to note, however, that since the many different squark masses can vary independently in the pMSSM and 
that the LHC constraints on, e.g., the $u_L,d_L$, $u_R$ and $d_R$ squarks are quite different, both SD and SI interactions may have substantial isospin-dependent 
contributions  so that $\sigma_p$ and $\sigma_n$ can be significantly different.}} The $Z$-exchange graph is sensitive to the Higgsino content of the LSP whereas 
the Higgs exchange graph probes the product of the LSP's gaugino and Higgsino content. Similarly, (valence) squark exchange is sensitive to the LSP's 
wino and bino content. 

\begin{figure}[htbp]
\centerline{\includegraphics[width=3.5in]{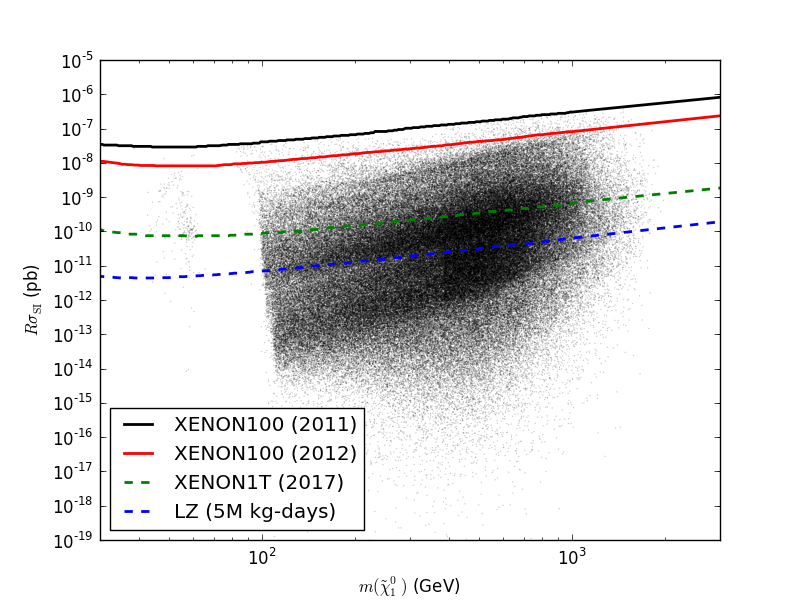}
\hspace{-0.50cm}
\includegraphics[width=3.5in]{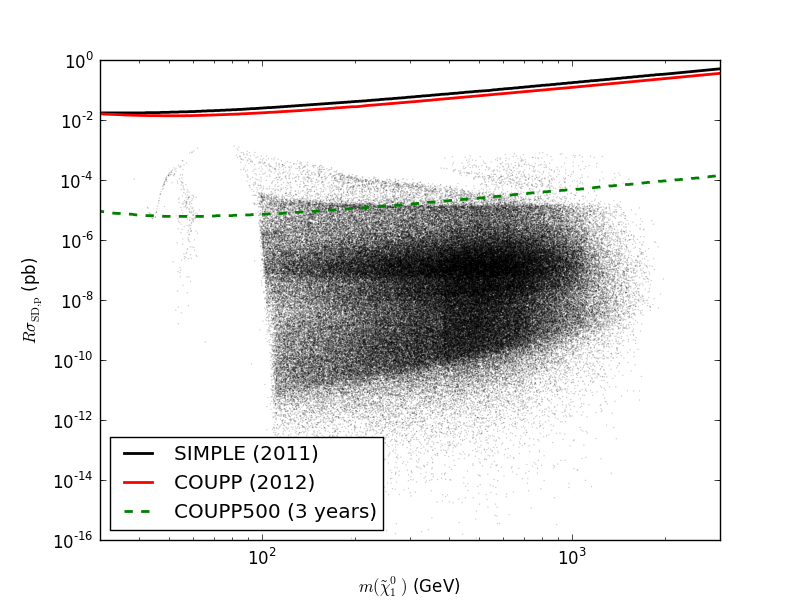}}
\vspace*{-0.10cm}
\caption{Scaled spin-independent (left) and spin-dependent (right) direct detection cross sections for our neutralino LSPs in comparison to current 
and future experimental sensitivities. The scaling factor accounts for the possibility that the calculated thermal relic density of the LSP is below 
that measured by WMAP.}
\label{fig3}
\end{figure}

Figure~\ref{fig3} displays the predicted experimentally observable SI and SD cross sections for our pMSSM model set together with several present and anticipated 
future experimental constraints. Note that the actual cross sections are appropriately scaled by the factor $R=\Omega h^2/\Omega h^2|_{WMAP}$ to account 
for the fact that most of our pMSSM models lead to a thermal relic density somewhat below that which saturates the WMAP value as discussed in the Introduction. 
There are several things to note in this Figure: ($i$) Future SI searches will cut rather deeply into the model set; a lack of signal at XENON1T (LZ) would 
exclude 23\% (50\%) of these models. {\it However}, this implies that about half of our models are not accessible to SI experiments due to their 
rather small scaled cross sections, $R\sigma$. Models with these values of $R\sigma$ tend to produce these small values due to both the suppression arising from 
their low thermal relic density as well as fact that the LSPs tend to be rather close to being pure weak eigenstates as discussed above. Note that if we only consider models which predict a relic density within 10\% of the critical density, the situation improves significantly, with 60\% (87\%) of models within the XENON1T (LZ) exclusion limit. ($ii$) SD measurements 
are still rather far from the pMSSM model predictions and we do not expect future SD measurements to have a significant impact on the parameter space; SD experiments such as COUPP500 
will only be able to exclude $\sim 2\%$ of the models in this set if no signal is seen.

While direct detection experiments have significant power to cover much of the pMSSM, clearly such experiments will need to be `supplemented' if we want to 
discover or exclude the full range of neutralino LSPs in the pMSSM model space. In the next two sections, we describe constraints on the pMSSM from indirect detection and neutrino telescope experiments.

\section{Indirect Detection: \Fermi ~LAT and CTA}

Indirect Detection plays a critical role in searches for DM and, in
the case of null results, can lead to very strong constraints on the
pMSSM parameter space. As will be seen below, both \Fermi ~and CTA can
contribute in different regions of the pMSSM parameter space in the
future.  CTA, in particular, will be seen to be extremely powerful in
the search for heavy LSPs which are mostly Higgsino- or wino-like and
that predict thermal relic densities within an order of magnitude of
the WMAP/Planck value. \Fermi, ~on the other hand, will be seen to be
mostly sensitive to well-tempered neutralinos that are relatively
light.

The most promising DM targets for both \Fermi ~LAT and CTA are those
with both a high DM density and low astrophysical gamma-ray
foregrounds.  These criteria have motivated a number of Galactic and
extragalactic targets including the Galactic Center (GC), dwarf
spheroidal galaxies (dSphs), and galaxy clusters. The expected
gamma-ray signal for DM annihilations is proportional to the integral
of the square of the DM density along the line of sight to the source
($J$).  The determination of $J$ is most reliable for DM-dominated
objects such as dSphs and galaxy clusters in which the DM distribution
can be robustly measured.
In the Milky Way halo the uncertainty on the DM distribution rapidly
increases as one approaches the inner galaxy where baryons dominate
the gravitational potential.  Kinematic data at large scales constrain
the density of DM at the solar radius to 0.2--0.4~GeV~cm$^{-3}$
\cite{Catena:2009mf, Bovy:2012tw}.  

Under well-motivated models for the DM distribution in the Galactic
halo, the GC is expected to be the most intense DM source in the sky.
CDM simulations predict that the Galactic DM halo should have a
density profile with an inner cusp, $\rho \propto r^{-\gamma}$ with
$\gamma \simeq 1.0$.  For an extrapolation of the Galactic DM density
profile with $\rho(r) = \rho_{\odot}(r/r_\odot)^{-1}$, the expected DM
signal from the GC region is approximately two orders of magnitude
greater than dSphs or galaxy clusters.  However the GC is also the
region of the sky with the highest density of gamma-ray sources and
brightest diffuse gamma-ray emission produced from the interaction of
cosmic rays with the interstellar medium.  These foregrounds
significantly limit the sensitivity of the \Fermi-LAT in the inner
galaxy and complicate the interpretation of any observed signals.  For
our study of \Fermi-LAT we focus on the sensitivity to the DM signal
from dSphs which are predominantly at high galactic latitudes where
the astrophysical foregrounds are much weaker.  Above 50~GeV the
diffuse emission from the Galaxy is much less intense relative to
other backgrounds than in the \Fermi-LAT sensitivity band.  The inner
galaxy is thus the preferred target for CTA under the assumption that
the Galactic DM halo possesses an inner cusp with $\gamma \simeq 1.0$.

\subsection{\Fermi ~LAT}

Here we follow the procedure developed in Ackermann et al.~(henceforth A11) \cite{Ackermann:2011wa} and extended in Cotta et al.~\cite{Cotta:2011pm} 
to constrain 
the annihilation cross section, \sigv, for each pMSSM model using \Fermi-LAT observations of ten dwarf spheroidal galaxies (dSphs).  Our two-year 
$\gamma$-ray event 
sample is identical to that described in A11, accepting photons in the energy range from $200\,\MeV < E < 100\,\GeV$ within 10\degree of each dSph. 
In accord 
with A11, we use the LAT ScienceTools\footnote{http://fermi.gsfc.nasa.gov/ssc/data/analysis/software} version v9r20p0 and the P6\_V3\_DIFFUSE 
instrument response functions.\footnote{http://fermi.gsfc.nasa.gov/ssc/data/analysis/scitools/overview.html} J-factors and associated statistical 
uncertainties for the dSphs are taken from Table 1 of A11, where they were calculated using 
line-of-sight stellar velocities and the Jeans equation \cite{Ackermann:2011wa}. 
Similar to Cotta 
et al., we use DarkSUSY 5.0.5 \cite{Gondolo:2004sc} to model the $\gamma$-ray spectrum from the annihilation of each pMSSM LSP. DarkSUSY calculates 
the total 
$\gamma$-ray yield from annihilation, as well as the rates into each of 27 final state channels.

We calculate a joint likelihood to constrain the annihilation cross section of each pMSSM model given the LAT observations 
coincident with the ten dSphs.  
Following A11, 
we incorporated statistical uncertainties in the J-factors of the dSphs as nuisance parameters in our likelihood formulation~(see Equation 1 
and the associated 
discussion in A11). This likelihood formulation includes both the flux normalizations of background $\gamma$-ray sources (diffuse and point-like) 
and the 
associated dSph J-factors and statistical uncertainties. No significant $\gamma$-ray signal is detected from any of the dSphs when analyzed 
individually or 
jointly for any of the pMSSM models.

For each of the pMSSM models, we calculate the maximum annihilation cross section, \sigvm, consistent with the null detection in the LAT data. 
We incorporate 
nuisance parameters to obtain a 95\% one-sided confidence interval on the value of \sigv using the profile likelihood method~\cite{Rolke:2004mj}. 
This 
one-sided 95\% confidence limit on \sigv serves as our value of \sigvm, which is compared to the true value of the annihilation cross section 
for each pMSSM 
model. We define the ``boost'' necessary to constrain a model as the ratio $\sigvm/\sigv$.

While the LAT data do not presently constrain any of the pMSSM models, it is useful to estimate the improvements expected over a 10 year mission 
lifetime. 
In the low-energy, background dominated regime, the LAT point source sensitivity increases as roughly the square-root of the integration time. 
However, in 
the high-energy, limited background regime (where many pMSSM models contribute), the LAT sensitivity increases more linearly with integration time. 
Thus, 
10 years of data could provide a factor of $\sqrt{5}$ to 5 increase in sensitivity. Additionally, optical surveys such as 
Pan-STARRS~\cite{Kaiser:2002zz}, 
the Dark Energy Survey~\cite{Abbott:2005bi}, and LSST~\cite{Ivezic:2008fe} could provide a factor of 3 increase in the number of Milky Way dSphs 
corresponding 
to an increased constraining power of $\sqrt{3}$ to 3~\cite{Tollerud:2008ze}.  Ongoing improvements in LAT event reconstruction, a better 
understanding of 
background contamination, and an increased energy range are all expected to provide additional increases in the LAT sensitivity. Thus, we find 
it plausible 
that the LAT constraints could improve by a factor of 10 compared to current constraints. 

\begin{figure}[htbp]
\centerline{\includegraphics[width=3.5in]{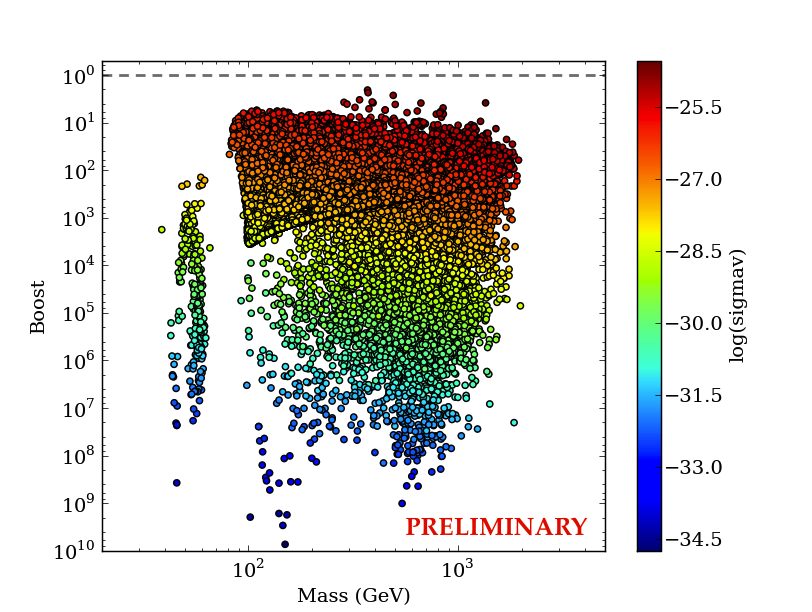}
\hspace{-0.50cm}
\includegraphics[width=3.5in]{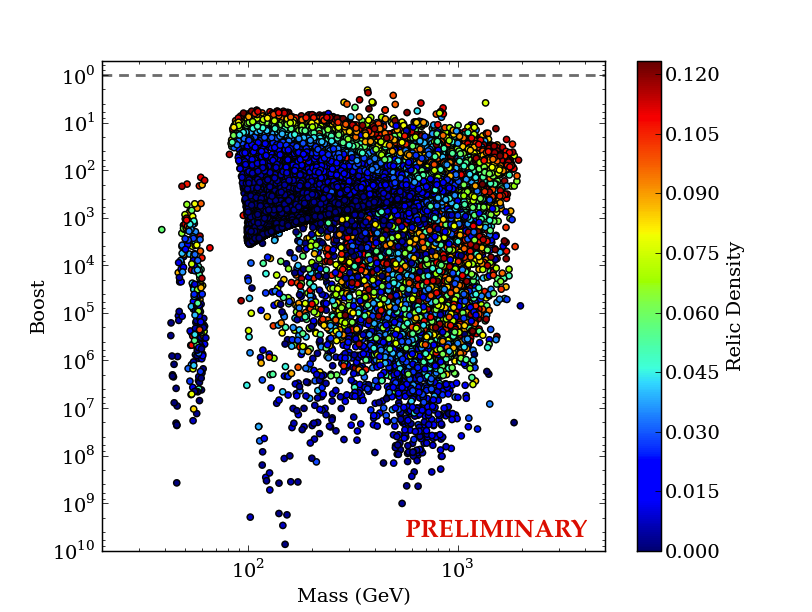}}
\vspace*{-0.10cm}
\caption{(Left) \Fermi-LAT boost factor vs. LSP mass for the pMSSM model set. The true cross section, \sigv, for each model is plotted on the 
color scale. (Right) Here the corresponding relic density for each model is plotted on the color scale.}
\label{fig4}
\end{figure}

In Figure~\ref{fig4} we display the boost required to constrain the various pMSSM models at $95\%$ CL based on the \Fermi-LAT dwarf analysis 
employing only the 
first 2 years of data colored-coded by either the annihilation cross section or the LSP thermal relic density. Here we see that the LAT analysis does not currently constrain any of our pMSSM models. However, as discussed above with more dwarfs and longer integration times we would expect an $\sim 10$-fold improvement in the 
sensitivity and thus 
all models with boost factors less than 10 would become accessible. We will assume this $\sim 10$-fold improvement in sensitivity for the analysis that follows.

\subsection{CTA}

The Cherenkov Telescope Array (CTA) \cite{Consortium:2010bc} is a
future ground-based gamma-ray observatory that will have sensitivity
over the energy range from a few tens of GeV to a few hundreds of TeV.
To achieve the best sensitivity over this wide energy range CTA will
include three telescope types: Large Size Telescope (LST, 23 m
diameter), Medium Size Telescope (MST, 10-12 m) and Small Size
Telescope (SST, 4-6 m).  Over this energy range the point-source
sensitivity of CTA will be at least one order of magnitude better than
current generation imaging atmospheric Cherenkov telescopes such as
H.E.S.S., MAGIC, and VERITAS.  CTA will also have an angular
resolution at least 2--3 times better than current ground-based
instruments, improving with energy from 0.1$^\circ$ at 100 GeV to
better than 0.03$^\circ$ at energies above 1 TeV.

The optimal DM search region for CTA will be limited by the CTA FoV of
8$^\circ$ to the area within $2^\circ$ to $3^\circ$ of the GC.  On these angular
scales the DM signal is predominantly determined by the DM
distribution in the inner galaxy ($R_{GC} <1$~kpc).  We model the
Galactic DM distribution with an NFW profile with a scale radius of 20
kpc normalized to 0.4~GeV~cm$^{-3}$ at the solar radius.  This model
is consistent with all current observational constraints on the
Galactic DM halo and represents a conservative expectation for the
inner DM profile in the absence of baryons.  Because the annihilation
signal is proportional to the square of the DM density, the projected
limits for CTA depend strongly on the assumptions that are made on the
shape and normalization of the Galactic DM halo profile.  The
projected limits presented here could change by as much as a factor of
10 given these uncertainties.

The potential of CTA for DM searches and testing other exotic physics
has been studied in detail by \cite{Doro:2012xx} using the projected
performance of several alternative array configurations
\cite{Bernlohr:2012we} with 18--37 MSTs and different combinations of
SSTs and LSTs.  For this study we use the projected sensitivity of a
candidate CTA configuration with 61 MSTs on a regular grid with 120~m
spacing \cite{Jogler:2012}.  This telescope layout is similar to the
anticipated layout of the baseline MST array with 18--25 telescopes
with an additional US contribution of 36 MSTs.  This configuration has
a gamma-ray angular resolution that can be parameterized as a function
of energy as $\theta \simeq 0.07^\circ(E/100$~GeV)$^{-0.5}$ and a
total effective area above 100 GeV of $\sim$10$^{6}$ m$^{2}$.  We
define the GC signal region as an annulus centered on the GC that
extends from 0.3$^\circ$--1.0$^\circ$ and calculate the sensitivity of
CTA for an integrated exposure of 500 hours that is uniform over the
whole region.  An energy-dependent model for the background in the
signal region is taken from a simulation of residual hadronic
contamination.  The uncertainty in the background model is calculated
for a control region with no signal contamination and a solid angle
equal to five times the signal region (14.3~deg$^{2}$).

We compute the sensitivity of CTA using a two-dimensional likelihood
analysis that models the distribution of detected gamma-rays in
energy and angular offset from the GC.  Following the same
procedure as the \Fermi-LAT analysis, we compute the model boost factor
as the ratio of the model cross section with the 95\% C.L. upper
limit on the annihilation cross section of a DM model with the same
spectral shape.  For models with LSP masses below 1~TeV, the
sensitivity of CTA is dominated by events at low-energy ($E\lesssim
300$~GeV).

\begin{figure}
  
  \centering
  \includegraphics[width=0.49\textwidth]{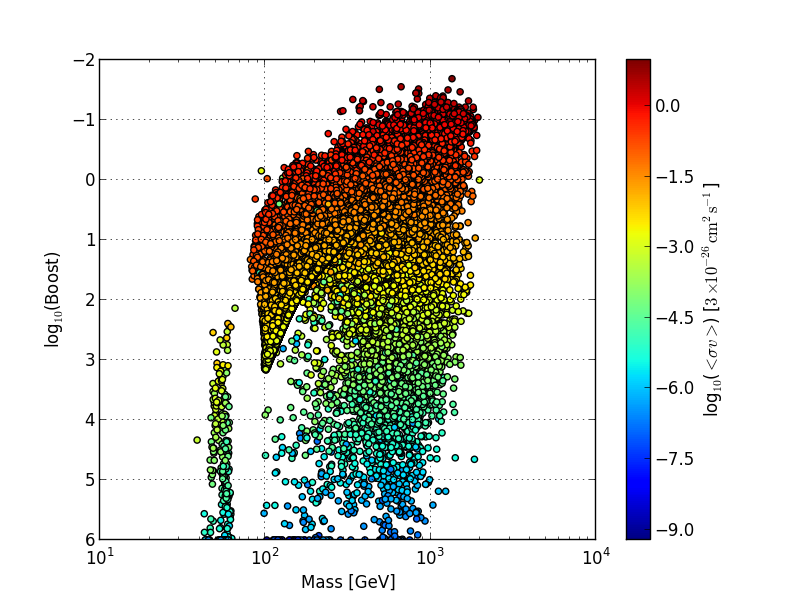}
  \includegraphics[width=0.49\textwidth]{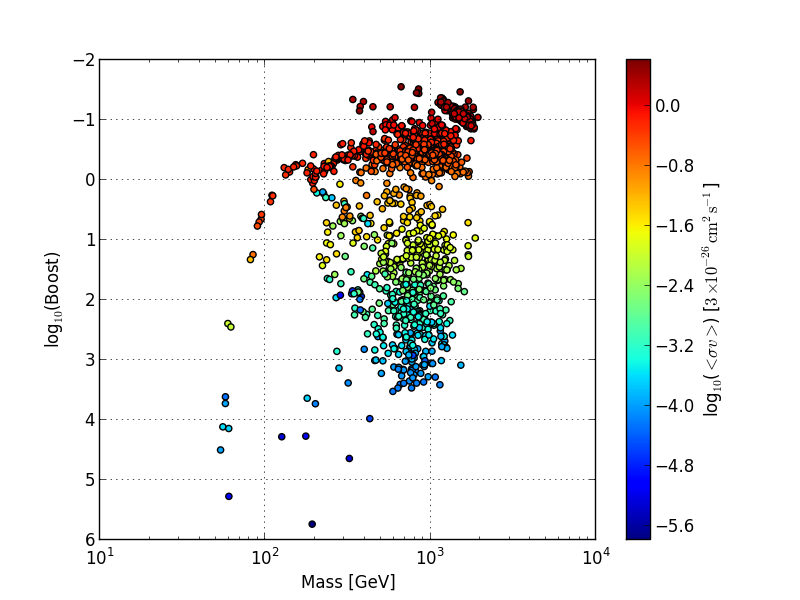}
  \caption{\label{fig:cta_pmssm_sigmav}CTA boost factor vs. LSP mass for the full pMSSM model set
    (left) and pMSSM models satisfying current experimental
    constraints and a LSP relic density consistent with the WMAP7 DM
    density. The true cross section, \sigv, for each model is plotted
    on the color scale.}
\end{figure}

Figure \ref{fig:cta_pmssm_sigmav} shows the distribution of the CTA
boost factor versus LSP mass for all pMSSM models and the subset of
models that have an LSP relic density consistent with 100\% of the DM
relic density.  CTA can exclude $\sim$20\% of the total model set and
$>$50\% of the models in the subset with the DM relic density.  Here
we see that $\sim 19\%$ of the models would be excluded by CTA if no
signal were to be observed.

\section{IceCube}

Neutralino dark matter can be captured and accumulated in the sun. Neutralinos in this relatively overdense population would then also sink to the 
solar core and annihilate. If the product of capture and annihilation cross sections is large enough this process leads to an equilibrium
 population of captured neutralinos whose annihilations are proportional to their \emph{elastic scattering cross-sections} 
\cite{Press:1985ug,Gould:1987ir} and that may be 
detectable by observing an excess of high-energy ($\geq \rm{GeV}$) solar neutrinos in $km^3$-scale neutrino telescope experiments~\cite{Silverwood:2012tp}.

Here we present predictions for an IceCube/DeepCore (IC/DC) search for neutralino DM in our model set. Our analysis closely 
follows that presented in \cite{Cotta:2011ht}. In
the results presented here we assume that each neutralino's relic density is given by the usual thermal calculation. We use 
DarkSUSY 5.0.6 \cite{Gondolo:2004sc} to simulate the 
(yearly-average) signal $\nu_{\mu}$/$\bar{\nu}_{\mu}$ neutrino flux spectra incident at the detector's position and convolve with 
preliminary
 $\nu_{\mu}$/$\bar{\nu}_{\mu}$ effective areas for muon events contained in DeepCore\footnote{These are the same effective areas that 
were used in \cite{Cotta:2011ht},
 referred to there as ``SMT8/SMT4."}. We consider a data set that includes $\sim 5$yr of data that is taken during austral winters 
(the part of the year for
 which the sun is in the northern hemisphere) over a total period of $\sim 10$yrs\;\footnote{In practice the IC/DC treatment of data 
is more sophisticated, classifying
events as through-going, contained and strongly-contained, and allowing for some contribution from data taken in the austral summer. 
We expect that inclusion of this data
 would affect our results at a quantitative, but not qualitative, level.}. An irreducible background rate of $\sim 10$ events/yr is 
expected from
cosmic ray interactions with nuclei in the sun. Here we will take (as discussed at greater length in \cite{Cotta:2011ht}) a detected 
flux of $\Phi=40$ events/yr as a conservative
 criterion for exclusion. 

\begin{figure}[hbtp]
    \centering
    \includegraphics[width=1.0\textwidth]{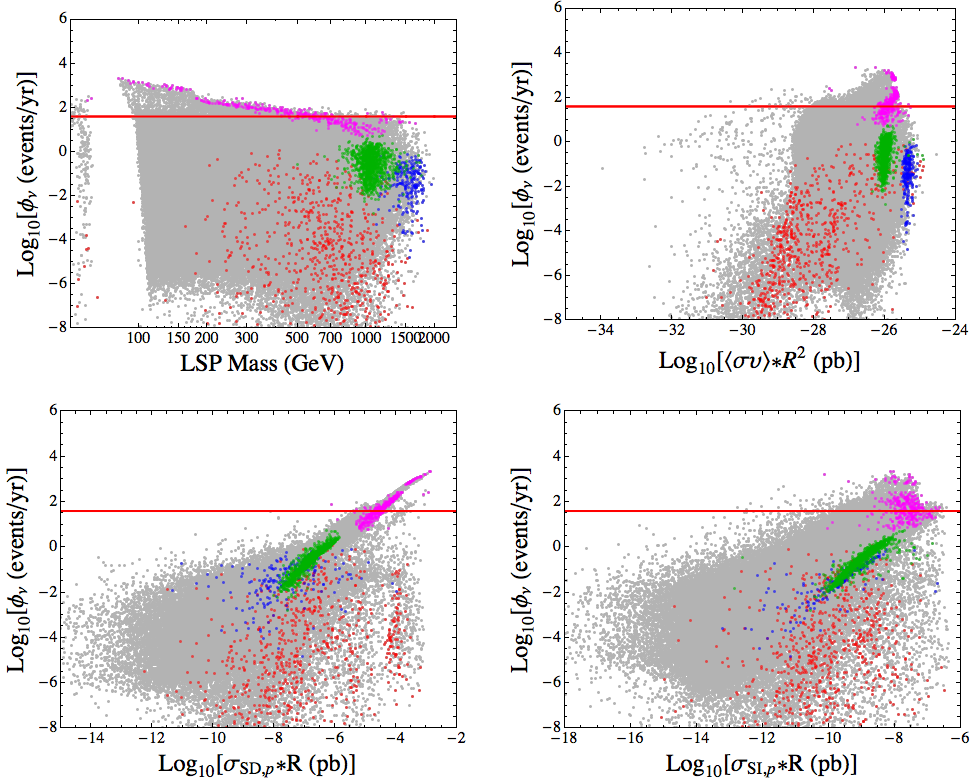}
    \caption{IC/DC signal event rates as a function of LSP mass (upper-left), thermal annihilation cross-section 
      $\langle\sigma\upsilon\rangle R^2$ (upper-right) and thermal elastic scattering cross-sections 
      $\sigma_{SD,p}$ and $\sigma_{SI,p}$ (lower panels). In all panels the gray points represent generic
      models in our full pMSSM model set, while WMAP-saturating models with mostly bino, wino, Higgsino 
      or mixed ($\leq$80\% of each) LSPs in are highlighted in red, blue, green and magenta, respectively.
      The red line denotes a detected flux of $40$ events/yr, our conservative estimate for exclusion.}
    \label{icdc}
  \end{figure}

The basic results of this analysis are presented in Figure \ref{icdc}. In Figure \ref{icdc}, we show all pMSSM models in our set using 
gray points, while 
highlighting WMAP-saturating models with mostly bino, wino, Higgsino or mixed ($\leq$80\% of each) LSPs in red, blue, green and magenta,
 respectively. Detectability is tightly correlated with the elastic scattering cross-sections ($\sigma_{SI}$ and $\sigma_{SD}$) while having
 little correlation with the annihilation cross-section $\langle\sigma\upsilon\rangle$, as expected. 

The biggest difference between these results and those of the previous analysis \cite{Cotta:2011ht}, which used a set of pMSSM models that
 were chosen to have relatively light ($\leq 1\tev$) sparticles, is that a much smaller percentage of the current pMSSM models are able to reach
 capture/annihilation equilibrium in the sun. This is due to the fact that so many of these models are nearly pure wino or Higgsino gauge 
eigenstates (which have both low
relic density and small capture cross-sections) and that the LSPs in this model set tend to be much heavier than those in the previous set. 
If one defines
out-of-equilibrium models as those with solar annihilation rates less than 90\% of their capture rates, we find that no such models can be 
excluded by IC/DC. In contrast,
 relatively light LSPs composed of a mixture of gaugino and Higgsino eigenstates have large scattering and annihilation cross sections and 
are highly detectable by IC/DC.
 We observe that all such WMAP-saturating well-tempered neutralinos with masses $m_{LSP}\leq 500\gev$ should be excluded by the 
IC/DC search (\emph{c.f.}, the magenta
 points in Fig. \ref{icdc}).

\section{Complementarity: Putting It All Together}

Now that we have provided an overview of the various pieces of data that go into our analysis, we can put them together to see what they (will) 
tell us about the nature of the neutralino LSP as DM~\cite{Boehm:2013qva} and, more generally, the pMSSM itself. We remind the reader that this is an ongoing analysis 
and that several future updates will be made to what we present here before completion. Since we only have 14 TeV results for the $\sim$ 30.7k neutralino LSP models that survive the 7+8 TeV searches and have $m_h=126\pm 3$ GeV (because of CPU limitations as described above), the main results presented below will use only the 7+8 TeV LHC searches listed in Table~\ref{SearchList}. We will also present some indicative results showing the sensitivity of the combined 7,8, and 14 TeV LHC analyses for the subset of neutralino LSP models with $m_h=126\pm 3$ GeV.

\begin{figure}[htbp]
\centerline{\includegraphics[width=3.5in]{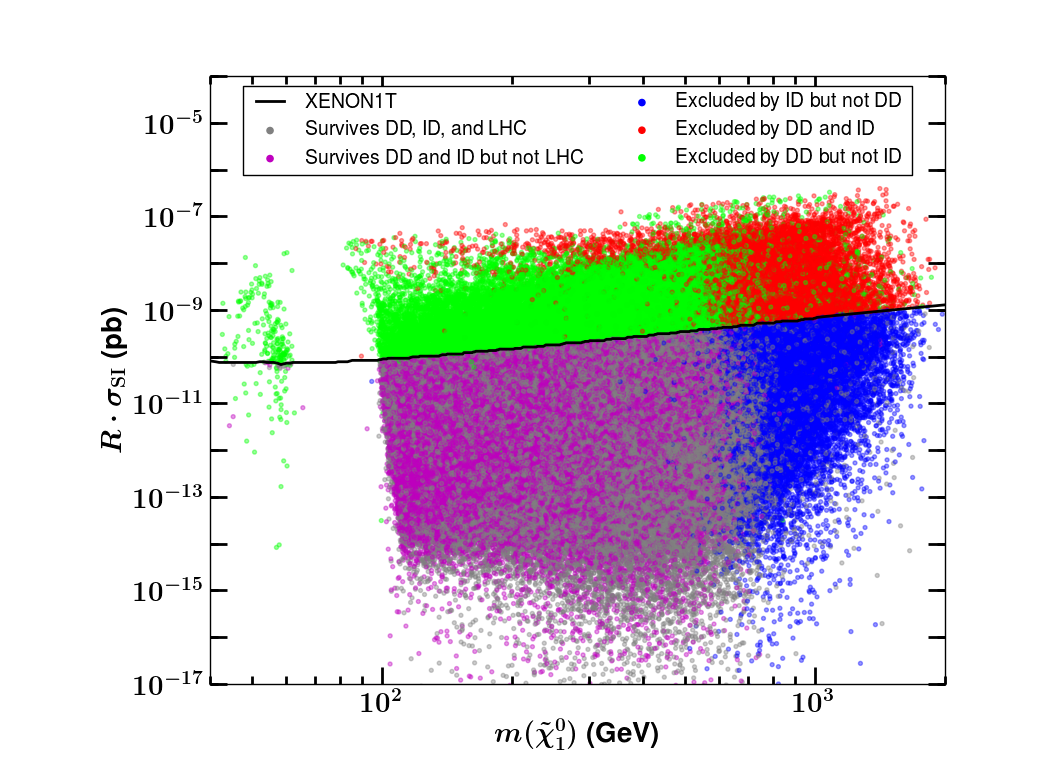}
\hspace{-0.50cm}
\includegraphics[width=3.5in]{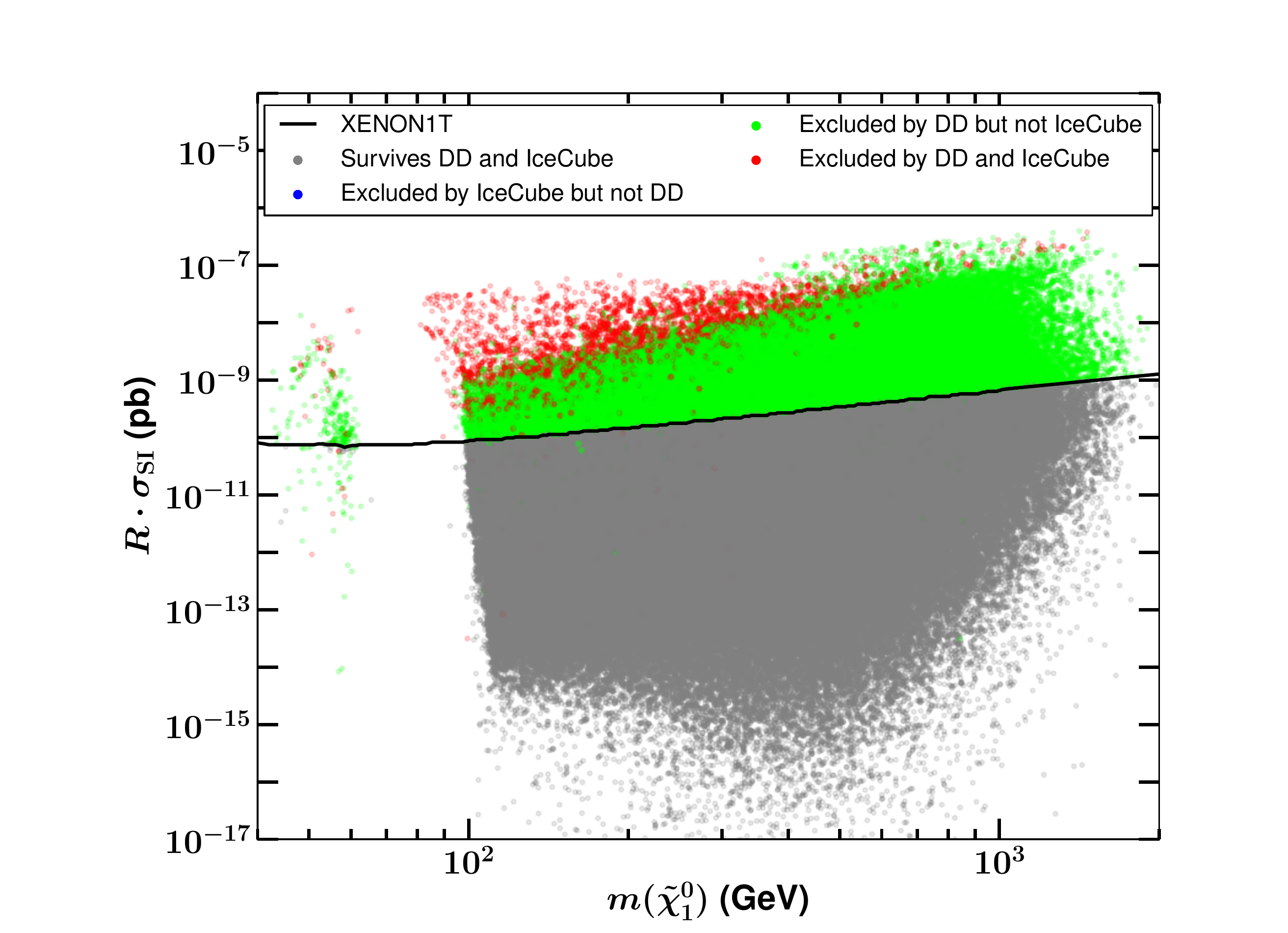}}
\vspace*{0.50cm}
\centerline{\includegraphics[width=3.5in]{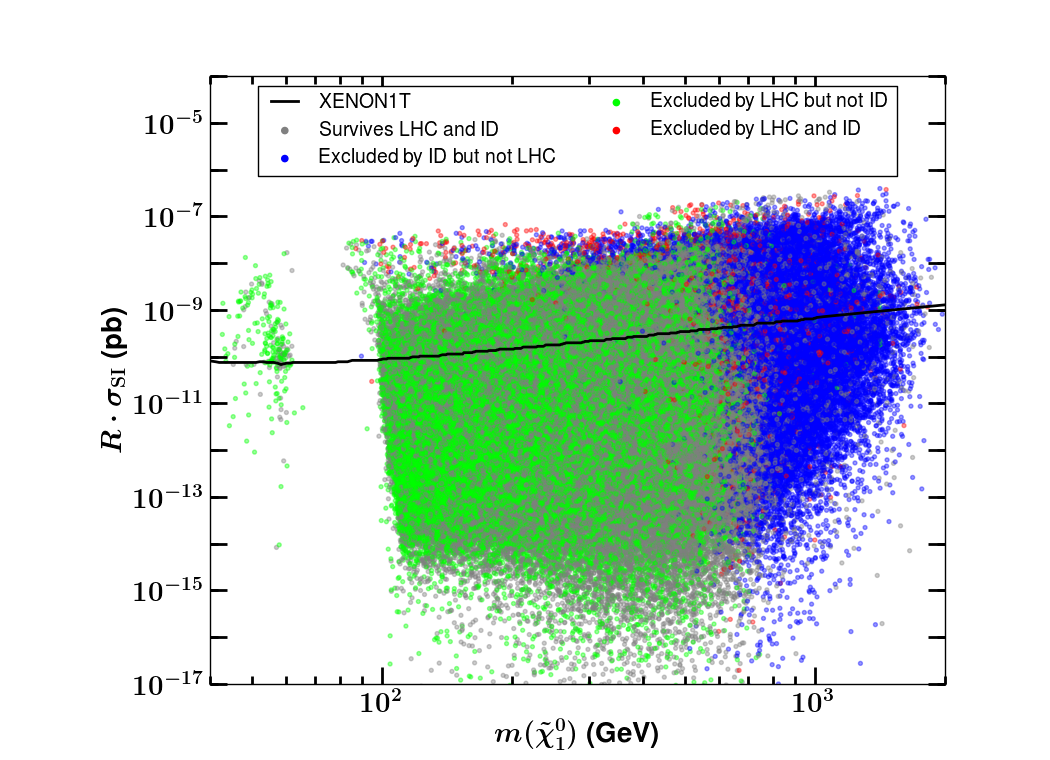}
\hspace{-0.50cm}
\includegraphics[width=3.5in]{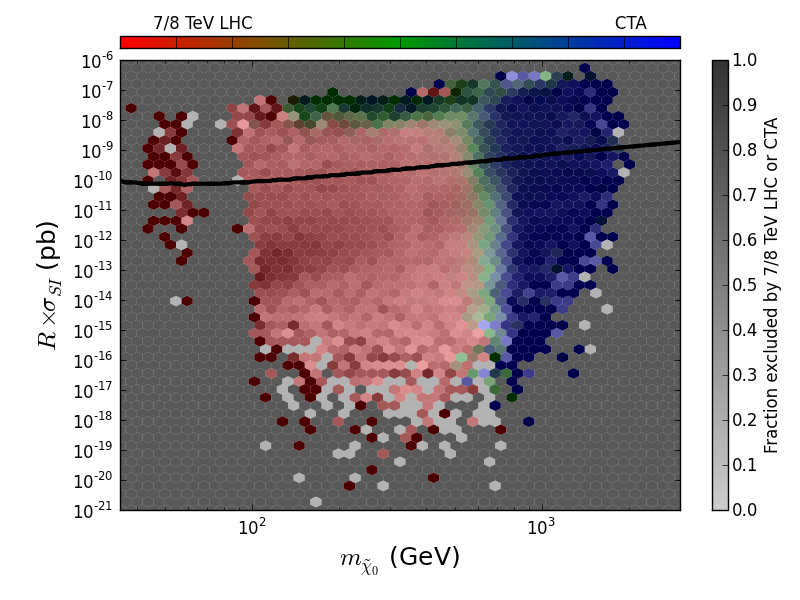}}
\vspace*{-0.10cm}
\caption{Comparisons of the models surviving or being excluded by the various searches in the LSP mass-scaled SI cross section plane as discussed 
in the text. The SI XENON1T line is shown as a guide to the eye.}
\label{figxx}
\end{figure}

Fig.~\ref{figxx} shows the survival and exclusion rates resulting from the various searches and their combinations in the LSP mass-scaled SI cross 
section plane. In the upper left panel we compare these for the combined direct detection (DD = XENON1T + COUPP500) and indirect detection (ID = \Fermi ~+ CTA) DM searches. Here we see 
that 11\% (15\%) of the models are excluded by ID but not DD (excluded by DD but not ID) while 8\% are excluded by both searches. On the other hand, we also see that 66\% of the models 
survive both sets of DM searches; 44\% of this subset of models, in turn, are excluded by the LHC searches. Note that the DD- and ID-excluded regions are all relatively well separated in terms of mass and cross section although there is some overlap between the sets of models excluded by the different experiments. In particular we see that the ID 
searches (here almost entirely CTA) are covering the heavy LSP region even in cases where the SI cross section is very low and likely beyond the reach 
of any potential DD experiment. Similarly, in the upper right panel we see that 22\% (0\%) of the models are excluded uniquely  
by DD (IceCube) only while 1\% can be simultaneously excluded by both sets of searches and 77\% would be missed by either search set. In the lower 
left panel, ID and LHC searches are compared and we see that 17\% (35\%) of the models would be excluded only by the ID (LHC) searches. However, 
2\% (47\%) of the models are seen to be excluded by (or would survive) both searches. The strong complementarity between the LHC, CTA and XENON1T experiments 
is evident here as CTA probes the high LSP mass region very well where winos and Higgsinos dominate, XENON1T chops off the top of the distribution 
where the well-tempered neutralino LSP states dominate, and the LHC covers the relatively light LSP region (independent of LSP type) rather well. Of course the 
strength of the LHC coverage will significantly improve when 14 TeV searches are included, as we will see below. In the lower 
right panel the relative contributions arising from the LHC and CTA searches to the model survival/exclusion are shown. Here the intensity of a given 
bin indicates the fraction of models excluded there by the combination of both CTA and the LHC while the color itself indicates the weighting from 
CTA (blue) and LHC (red). It is again quite clear that CTA completely dominates for large LSP masses (which correspond to the mostly wino and Higgsino 
LSPs) and also competes with the LHC throughout the band along the top of the distribution where the LSP thermal relic density is approximately 
saturating WMAP/Planck. The LHC covers the rest of the region but is not yet as effective as CTA (excluding a smaller fraction of models in the relevant 
bins).

\begin{figure}[htbp]
\centerline{\includegraphics[width=3.5in]{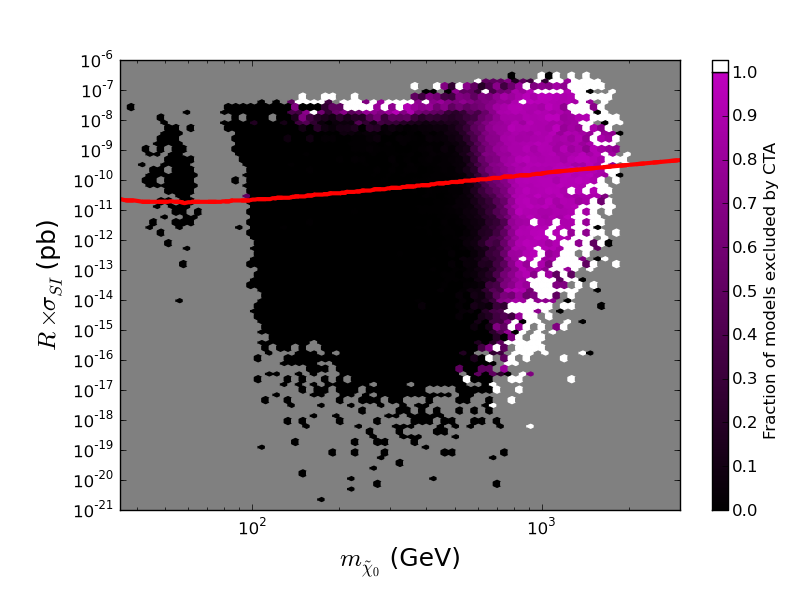}
\hspace{-0.50cm}
\includegraphics[width=3.5in]{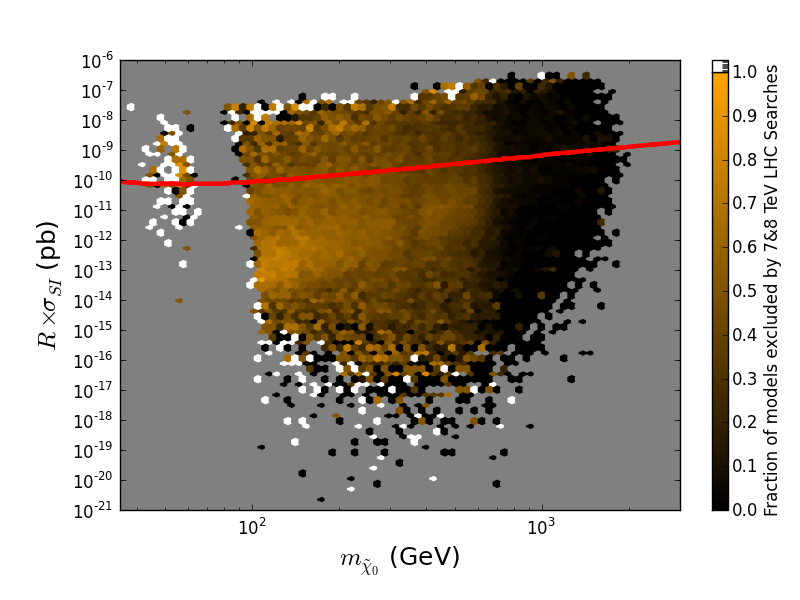}}
\vspace*{0.50cm}
\centerline{\includegraphics[width=3.5in]{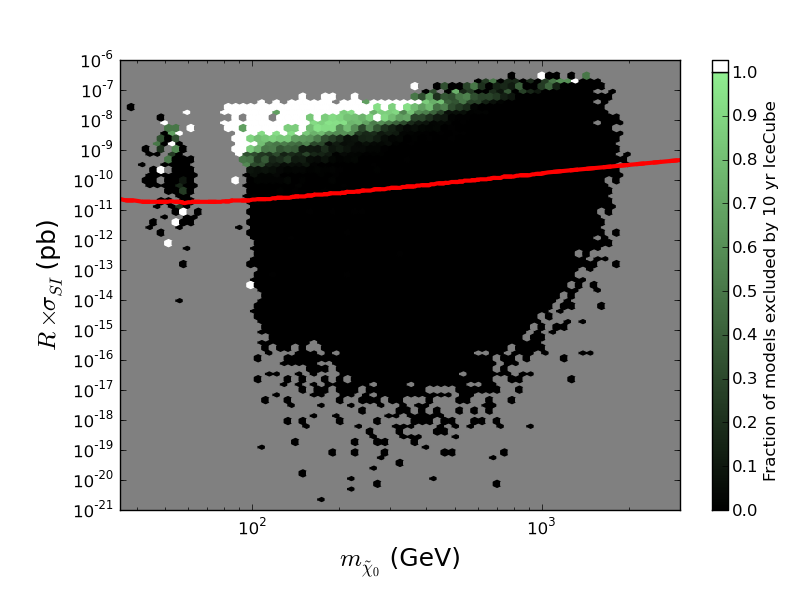}
\hspace{-0.50cm}
\includegraphics[width=3.5in]{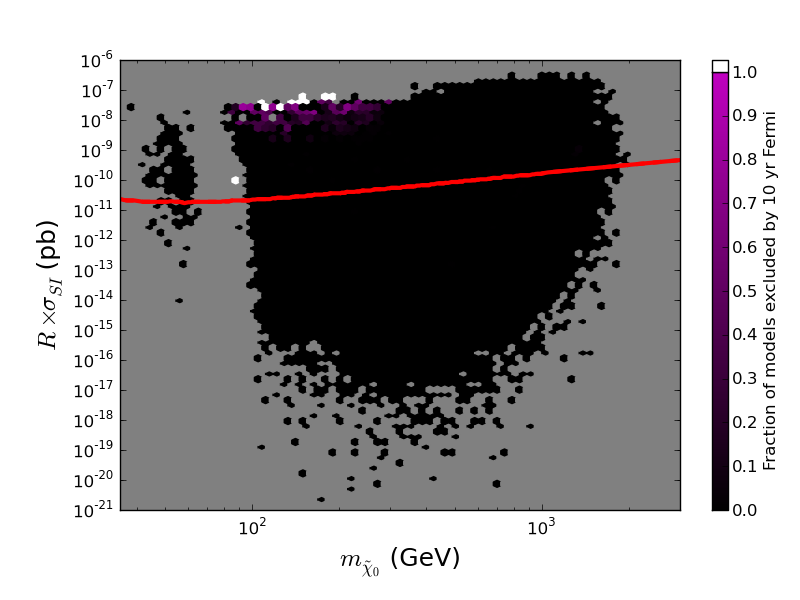}}
\vspace*{-0.10cm}
\caption{Comparisons of the various search capabilities in the LSP mass-scaled SI cross section plane as probed by the fraction of models excluded 
by CTA (top left), the LHC (top right), IceCube (bottom left) and \Fermi ~(bottom right). The potential SI bound from XENON1T is also shown in all 
cases.}
\label{figyy1}
\end{figure}

In order to study the complementarity arising from the various searches in more detail, we found that it is very useful to project their 
individual capabilities onto several sets of fixed parameter planes that are directly related to one of the search categories. We begin these projections 
with Fig.~\ref{figyy1} which compares the various search capabilities in the familiar LSP mass-scaled SI cross section plane, most relevant for the SI DD 
searches, as defined by the fraction of models excluded in the various bins. Here we see the contributions to the search coverage coming uniquely 
from CTA (top left), the LHC (top right), IceCube (bottom left) and \Fermi ~(bottom right) and also show the limit that could arise from XENON1T 
for comparison. Here we see that both IceCube and \Fermi ~probe models with low LSP masses and large SI cross sections, where the LSP tends to be a 
bino-Higgsino admixture; this region is also accessible to the DD experiments such as XENON1T. On the other hand, CTA has access to the heavy LSP region, 
where there are a large fraction of relatively pure winos and Higgsinos, while the LHC coverage is concentrated (for now) on the relatively low mass 
LSP region. Interestingly, the LHC searches are not independent of the SI cross-section; a region of enhanced exclusion fraction is seen for SI cross sections near $\sim 10^{-13}$ pb. Models with cross sections in this region mostly have wino-like LSPs with light squarks; wino-like LSPs with heavier squarks have a lower SI cross-section, while Higgsinos and mixed states tend to have a higher SI cross-section whether or not light squarks are present.

\begin{figure}[htbp]
\centerline{\includegraphics[width=3.5in]{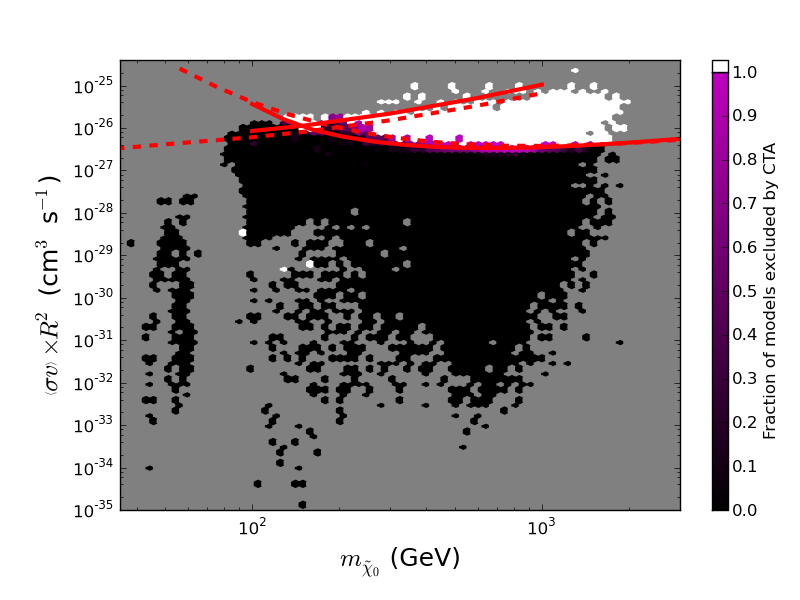}
\hspace{-0.50cm}
\includegraphics[width=3.5in]{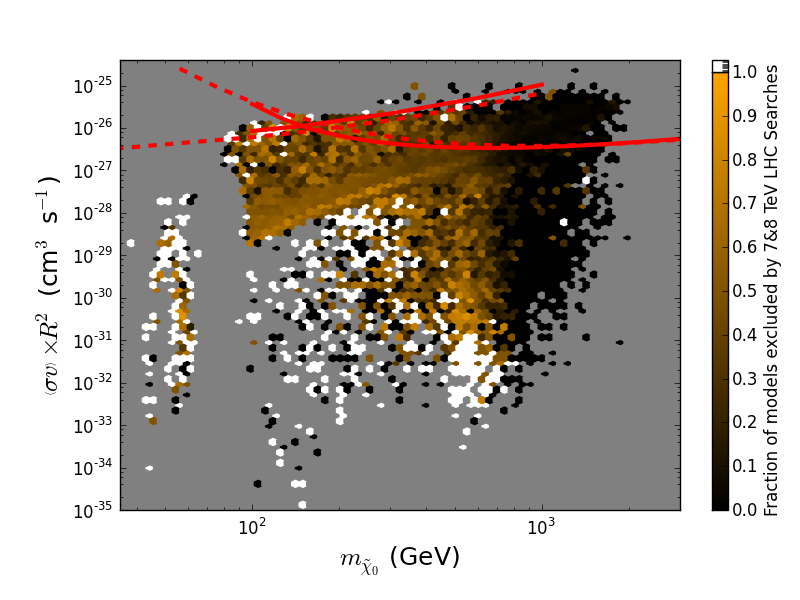}}
\vspace*{0.50cm}
\centerline{\includegraphics[width=3.5in]{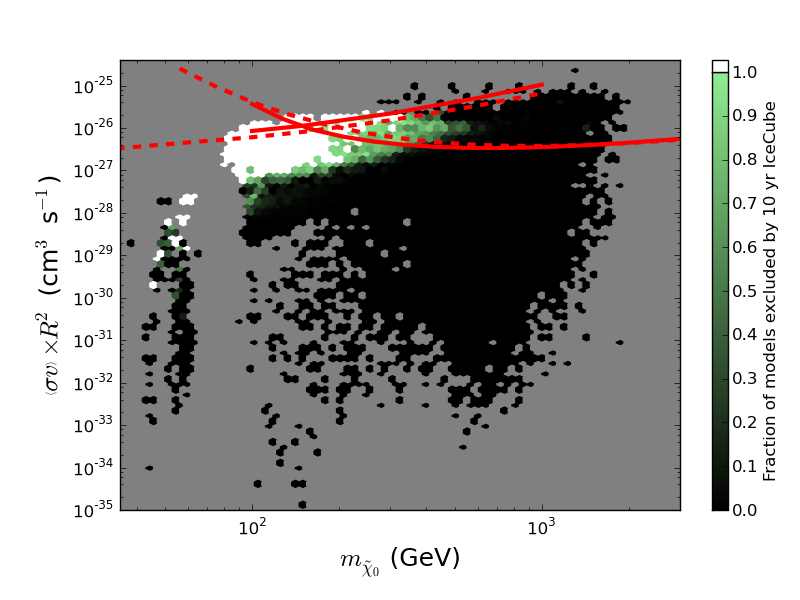}
\hspace{-0.50cm}
\includegraphics[width=3.5in]{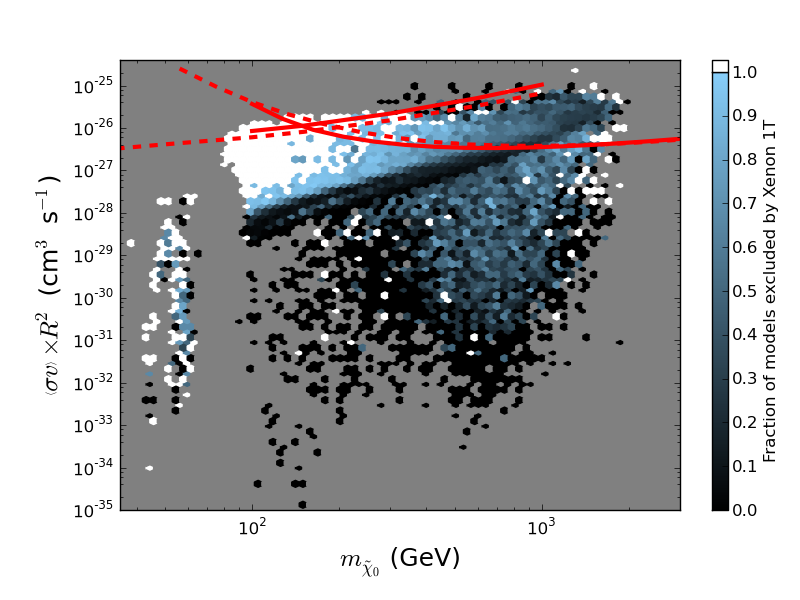}}
\vspace*{-0.10cm}
\caption{Comparisons of the various search capabilities in the LSP mass-LSP pair annihilation cross section plane, shown by the fraction of 
models in each bin which are excluded by CTA (top left), the LHC (top right), IceCube (bottom left) and XENON1T (bottom right). The potential bounds from \Fermi ~and 
CTA are also shown in all cases as curves through the upper left and right parts of the figures, respectively. The dashed (solid) curves are 
for $100\%$ annihilation into $b\bar b~(W^+W^-)$ final states only.}
\label{figyy2}
\end{figure}

Taking these same results we can now project them in to the plane which is most relevant for DM ID searches. Fig.~\ref{figyy2} compares the various 
search capabilities as above but now in the LSP mass-LSP pair annihilation cross section plane, for which both \Fermi ~and CTA are directly applicable. Here 
we see the fraction of models that would be excluded by searches at CTA (top left), the LHC (top right), IceCube (bottom left) 
and XENON1T (bottom right). The potential bounds from \Fermi ~and CTA are also shown in all cases as curves penetrating through the upper left and right 
parts of the figures, respectively. Here the dashed (solid) curves correspond to the limit obtained when the LSP pair annihilation final state is exclusively $b\bar b~(W^+W^-)$; we emphasize that a generic LSP in a pMSSM model may annihilate to many different final states beyond these two cases. Note again that CTA is primarily sensitive to models with LSP mass above 100 GeV and with \sigv relatively close to  
the thermal relic value.  As in the previous figure, the LHC is mainly effective in the lower LSP mass region. In addition, the LHC searches are seen to be particularly efficient along a thin line starting at annihilation cross sections of $\sim 10^{-29}\,\frac{cm^3}{s}$ for 100 GeV LSPs and increasing with LSP mass; this line corresponds to models with a pure wino LSP, which are subject to strong constraints from searches for heavy stable charged particles at the LHC. (Note that this line is below the projected reach of the ID searches for LSP masses accessible to the current LHC searches.) IceCube is seen to nicely complement CTA in the region with large cross sections but low LSP 
masses. The projection of the XENON1T coverage onto this plane is interesting with most of the exclusion appearing at lower LSP masses where 
the LSP pair annihilation cross section is also large. However, there is also an extended, somewhat diffuse region of substantial coverage throughout 
the entire right half of the parameter space as well as on the $Z,h$ funnel `island' at small LSP masses.

\begin{figure}[htbp]
\centerline{\includegraphics[width=3.5in]{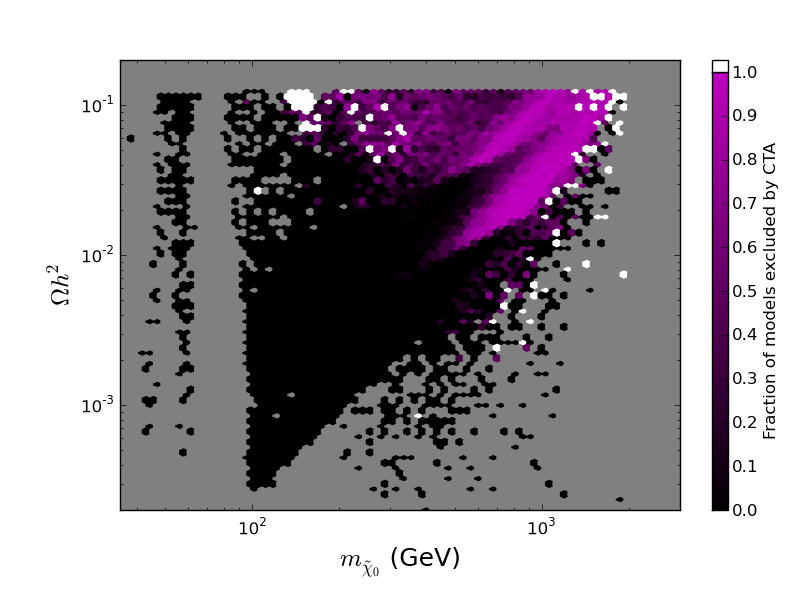}
\hspace{-0.50cm}
\includegraphics[width=3.5in]{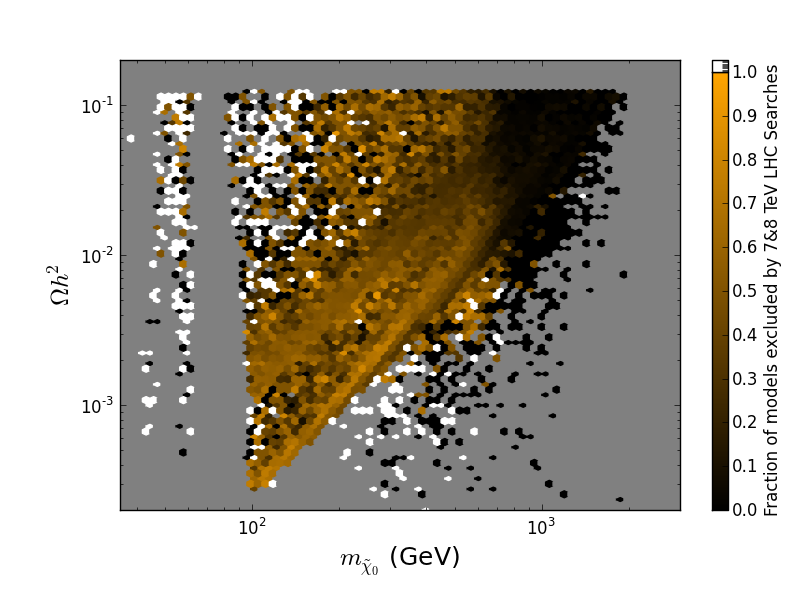}}
\vspace*{0.20cm}
\centerline{\includegraphics[width=3.5in]{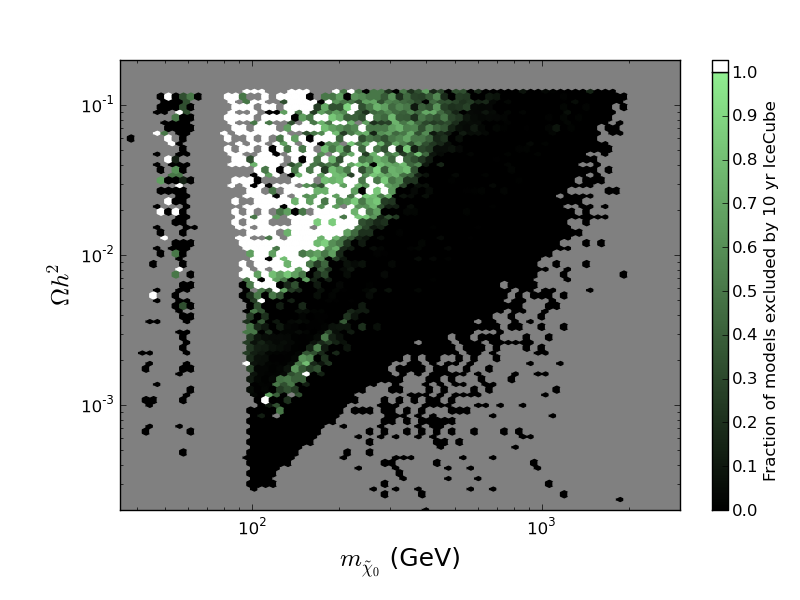}
\hspace{-0.50cm}
\includegraphics[width=3.5in]{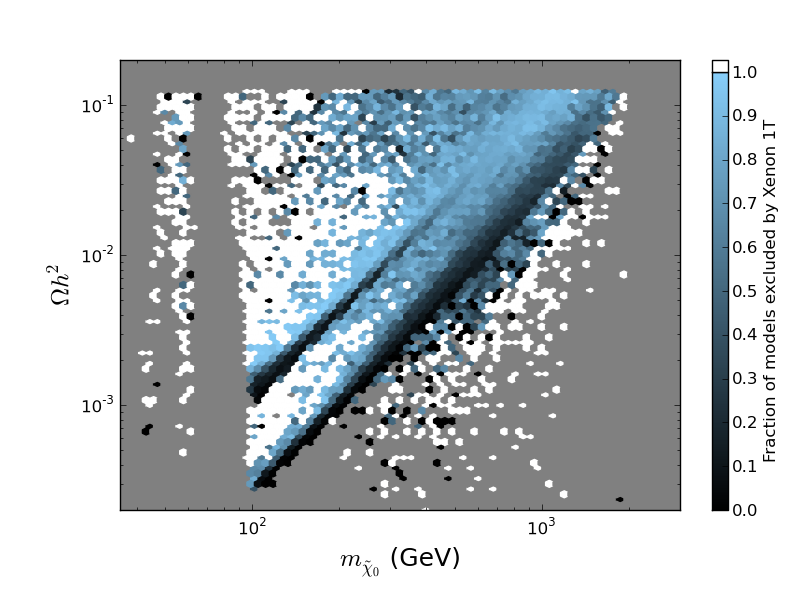}}
\vspace*{0.20cm}
\centerline{\includegraphics[width=3.5in]{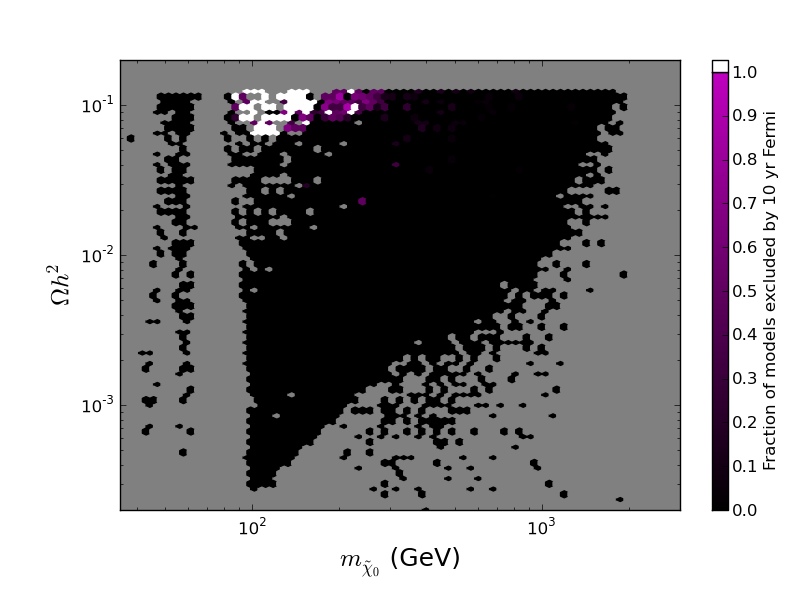}}
\vspace*{-0.10cm}
\caption{Comparisons of the various search capabilities in the LSP mass-relic density plane as probed by the fraction of models excluded by 
CTA (top left), the LHC (top right), IceCube (middle left), XENON1T (middle right) and \Fermi ~(bottom).}
\label{figyy3}
\end{figure}

As a final example of these projections, we now consider the thermal relic density-LSP mass plane; the relic density plays a `linear' role in the DD 
searches but a `quadratic' one in the ID case (and none at all as far as the LHC is concerned). In Fig.~\ref{figyy3} we display the coverage 
comparisons for the various experiments: CTA, LHC, IceCube, XENON1T and \Fermi, ~respectively. Again we see that multiple searches can make 
overlapping contributions in various parts of this parameter space plane{\footnote {In the case of an actual DM {\it discovery}, the existence of 
substantial overlapping regions in this parameter space will be of significant help in trying to determine the specific nature of the LSP.}}.  These 
plots show several things: First, CTA, as expected, does an excellent job at accessing most of the models with LSP masses above $\sim 250$ GeV that 
saturate the relic density. However, for larger masses the CTA coverage extends down to relic densities as much as a factor of $\sim 10$ or so below the 
WMAP/Planck value. \Fermi ~ is seen to cover only the low LSP mass region with relic densities not far from the thermal value while IceCube can go 
to much lower relic densities provided the LSP mass is below $\sim 500$ GeV or so. XENON1T displays coverage throughout this plane but does best for 
LSP masses below $\sim 300$ GeV, providing coverage even for models with very low relic densities. Of course, even for masses up to 1-2 TeV, 
XENON1T still provides quite decent model coverage in this parameter plane. As noted already, most of the impact of the LHC is at present seen to be at 
lower LSP masses below $\sim 500$ GeV. The LHC coverage is relatively uniform as far as the value of the relic density is concerned except in the 
case of very light LSPs where the coverage is very strong. Of course, we again remind the reader that extending the LHC energy to 14 TeV will substantially improve its effectiveness for excluding heavy LSPs.

\begin{figure}[htbp]
\centerline{\includegraphics[width=7.0in]{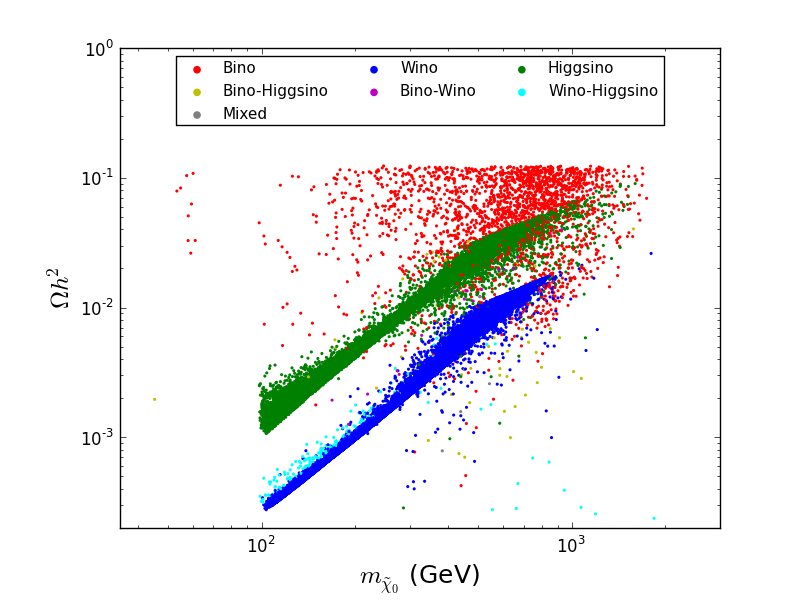}}
\vspace*{-0.10cm}
\caption{Thermal relic density as a function of the LSP mass for all pMSSM models, surviving after all searches, color-coded by the 
electroweak properties of the LSP. Compare with Fig.~\ref{fig00}. }
\label{figzz}
\end{figure}

Finally, Fig.~\ref{figzz} shows the impact of combining all of the different searches in this same $\Omega h^2$-LSP mass plane which should be compared 
with that for the original model set as generated that is shown in Fig.~\ref{fig00}. Here we see that ($i$) the models that were in the light $h$ and $Z$-funnel 
regions have essentially evaporated (and further measurements of the invisible width of Higgs as well as the standard SUSY searches will likely further restrict these), 
($ii$) the well-tempered neutralinos are now seen to be completely gone, mostly due to XENON1T and IceCube; ($iii$) the possibility of almost pure Higgsino or 
wino LSPs even approximately saturating the relic density has vanished thanks to CTA and ($iv)$ the mixed wino-Higgsino models, due to a combination of measurements, 
have also completely disappeared. ($v$) The only models remaining which {\it do} saturate the WMAP/Planck relic density are those with binos with resonant or co-annihilations. ($vi$) We 
find that $\sim 63\%$ of all the pMSSM models have been excluded by at least one of the searches considered in this paper.

\section{Complementarity with the 14 TeV LHC}

\begin{figure}[htbp]
\centerline{\includegraphics[width=3.5in]{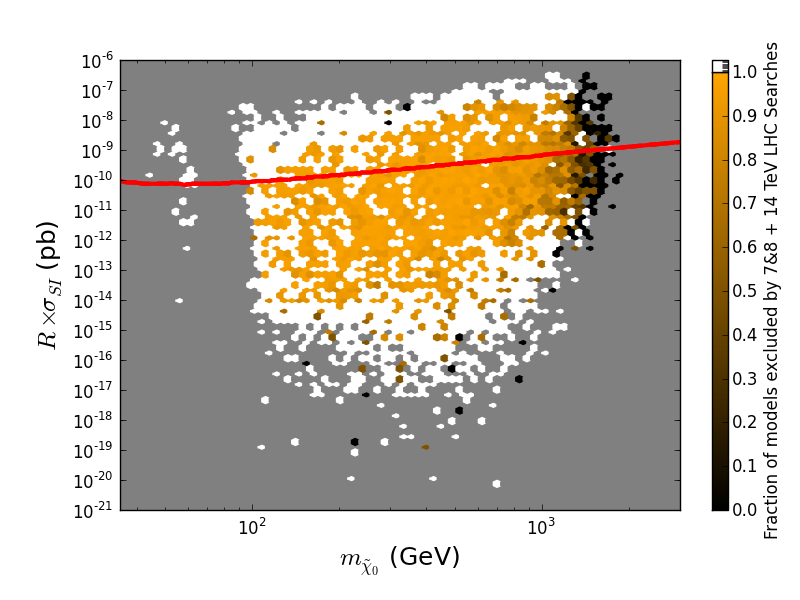}
\hspace{-0.50cm}
\includegraphics[width=3.5in]{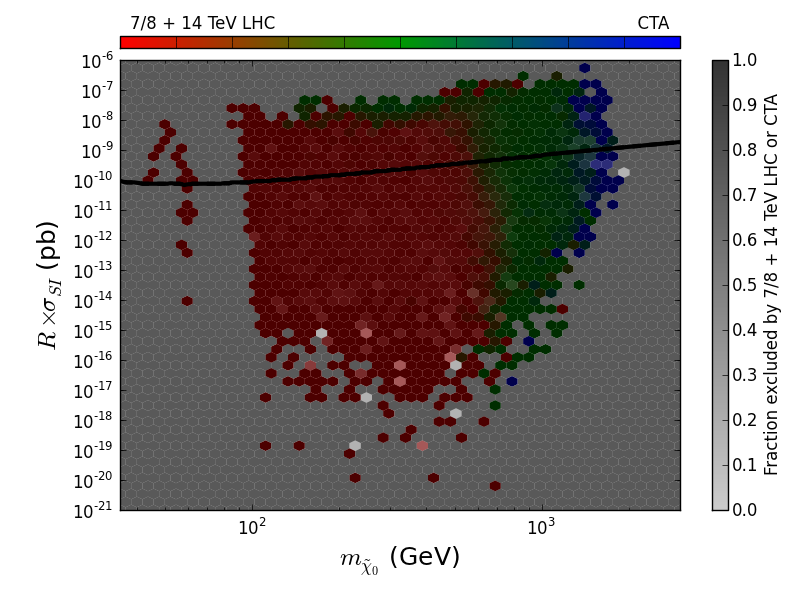}}
\vspace*{-0.10cm}
\caption{The fraction of models excluded by the 14 TeV Jets+MET search with 300 fb$^{-1}$ for models with the correct higgs mass, shown in the LSP mass-scaled SI cross section plane (left panel) and a comparison between the fractions excluded by the LHC and CTA in this plane (right panel). The potential SI bound from XENON1T is also shown in all 
cases.}
\label{14TeVComp}
\end{figure}

We now consider the effect of adding the 14 TeV jets + MET search with 300 fb$^{-1}$ of integrated luminosity to the full set of 7+8 TeV searches, considering only the subset of models with $m_h = 126\pm 3$ GeV. We note that both the LHC and dark matter results are essentially independent of the Higgs mass, so that the results for this subset should reproduce the results we would get for the full model set, albeit with lower statistics. The left panel of Figure~\ref{14TeVComp} shows the reach of the combined LHC searches in the LSP mass - SI cross-section plane for models with $m_h = 126\pm 3$ GeV. Comparing this figure to the upper right panel of Figure~\ref{figyy1}, we see two key changes with the inclusion of the 14 TeV search. First, in both cases the effectiveness of the LHC searches is seen to fall off sharply above a particular LSP mass, since above this limit the spectrum is generically either too heavy or too compressed to observe. Adding the 14 TeV searches effectively doubles this cutoff, from $\sim$ 700 GeV to $\sim$ 1400 GeV, so that most LSPs in our model set are now accessible given colored sparticle masses below $\sim$ 2-3 TeV. Second, we see that the LHC now excludes a very high fraction (but not all) of the models with LSPs lighter than the aforementioned cutoff. The large fraction of models which are excluded with 14 TeV data is unsurprising, since our chosen upper limits for the sparticle masses were designed to ensure that most models would be (at least kinematically) accessible at the 14 TeV LHC.

The right panel of Figure~\ref{14TeVComp} shows a comparison between the reaches of the LHC and CTA in this plane, analogous to the lower right panel of Figure~\ref{figxx}. We see that the LHC and CTA sensitivities now exhibit a sizable region of overlap. However, the blue region on the far right edge of the plot shows that CTA will still be sensitive to LSP masses beyond the reach of the LHC. We also note that LSPs heavy enough to be seen by CTA are generally too heavy to be seen in direct (e.g. monojet) searches, meaning that the LHC is excluding these models by observing other (mostly colored) sparticles. It is therefore important to note that CTA has the potential to exclude winos and Higgsinos with nearly thermal relic densities regardless of the characteristics of the rest of the sparticle spectrum. Of course, the improved reach of the LHC means that there is also an increasing overlap between the LHC and Xenon 1T, however we see that (as before) the LHC searches are mostly independent of the SI cross-section (the white areas around the edges of the plot generally result from low statistics in those regions). Of course, from a discovery perspective, the increased overlap means that there is more potential for a signal to be observed by two, or even all three, experiments, which would greatly aid in characterizing the LSP and model properties. Finally, we note that not only has the total fraction of models excluded by the combined experiments increased dramatically (from 63\% to 97.4\%) as a result of the improved LHC reach, but the fraction of models not seen by the LHC which are excluded by direct or indirect detection has increased (from 41\% to 49\%), since more of the undetected models have heavy LSPs and are therefore likely to be excluded by CTA.

\section{Conclusion}

The results presented here are preliminary but directly lead us to a number of interesting conclusions which are already quite clear and which we expect to strengthen 
in the future:

\begin{itemize}

\item{Even if the LSP {\it does not} make up all of the DM, it can still be observed in both direct and indirect detection experiments as well as in 
neutrino experiments such as IceCube. Of course, searches at the LHC are not influenced by the LSP relic density.}

\item{The set of models remaining after all the searches are performed that saturate the thermal relic density consist only of those with (co)annihilating bino LSPs.}

\item{SI direct detection, CTA, and the LHC do most of the heavy lifting in terms of complementary searches covering the pMSSM parameter space.}

\item{Multiple/overlapping searches allow for extensive parameter space coverage which will be of particular importance if a DM signal is observed.}

\item{Most of the experiments are seen to provide complementary probes of the pMSSM parameter space.} 

\item{The strength of the LHC component in these searches increases significantly with the inclusion of the 14 TeV LHC Jets + MET search. However, the Jets + MET search is dependent on the rest of the model spectrum and therefore does not provide complete coverage of any given LSP scenario, in contrast to DM searches which rely more directly on the LSP properties.}

\end{itemize}

In summary, the pMSSM provides an excellent tool for studying complementarity between different approaches to the search for dark matter.

\section*{Acknowledgments}

The authors would like to thank J. Feng, K. Matchev, and T. Tait for useful discussions. This work was supported by the Department of Energy, Contract DE-AC02-76SF00515, the Department of Energy Office of Science Graduate Fellowship Program (DOE SCGF), made possible in part by the American Recovery and Reinvestment Act of 2009, administered by ORISE-ORAU under contract no. DE-AC05-06OR23100, and the National Science Foundation under grant PHY-0970173.

\end{document}